\documentclass[a4paper,12pt]{article}
\usepackage{latexsym}
\usepackage{lmodern}
\usepackage{tikz}
\usetikzlibrary{shapes}
\usetikzlibrary{plotmarks}
\usepackage{amsmath,amsfonts,amssymb,amsthm,amstext,amscd,mathtools,eucal,graphicx,xcolor,esint,color}
\usepackage[all]{xy}
\usepackage{dsfont,epiolmec,comment}
\usepackage{slashed,ccaption,cite}
\usepackage{mathrsfs, calligra}
\usepackage{yfonts}
\usepackage{float}

\usepackage[unicode=true,
 bookmarks=false,
 breaklinks=false,pdfborder={0 0 1},backref=false,colorlinks=true]
 {hyperref}
\hypersetup{pdfauthor={ Shahin Sheikh-Jabbari and A.A. Abolhasani},
 linkcolor=blue,urlcolor=magenta,filecolor=green,citecolor=red,pdfstartview=FitV,bookmarksopen=true}
\usepackage{amsthm}
\usepackage{amsmath}
\usepackage{graphicx}
\usepackage{amssymb}
\usepackage{hyperref, comment}
\usepackage{subcaption}
\usepackage{simplewick}
\usepackage{soul}
\usepackage{mathrsfs}
\usepackage{axodraw2}

\makeatletter \@addtoreset{equation}{section}

\usetikzlibrary{decorations.pathmorphing}
\tikzset{snake it/.style={decorate, decoration=snake}}

\newcommand{\gsim}{\lower.7ex\hbox{$\;\stackrel{\textstyle>}{\sim}\;$}}
\newcommand{\lsim}{\lower.7ex\hbox{$\;\stackrel{\textstyle<}{\sim}\;$}}

\newcommand{\be}{\begin{equation}}
\newcommand{\ee}{\end{equation}}
\newcommand{\bea}{\begin{eqnarray}}
\newcommand{\eea}{\end{eqnarray}}

\newcommand{\expect}[1]{\left\langle #1 \right\rangle}

\newcommand{\bsb}{\boldsymbol}

\newcommand{\vphi}{\varphi}

\def\md{\mathrm{d}}

\def\q{{\bsb q}}
\def\p{{\bsb p}}
\def\k{{\bsb k}}

\def\K{{\bsb k}}
\def\x{{\bsb x}}

\def\y{{\bsb y}}

\usepackage{colortbl}
\definecolor{summersky}{cmyk}{0.71,0.33,0,0.14}
\definecolor{flamingo}{cmyk}{0,0.51,0.71,0.14}
\definecolor{rp}{cmyk}{0.2, 1, 0.6, 0}
\definecolor{pacificblue}{cmyk}{0.95,0.3,0, 0.19}
\definecolor{gray60}{cmyk}{0.4,0.4,0,0.8}
\definecolor{green94}{cmyk}{94,0,100,0}
\definecolor{green80}{cmyk}{80,0,90,0}

\definecolor{darkgreen}{rgb}{0,0.3,0}
\definecolor{darkblue}{rgb}{0,0,0.3}
\definecolor{darkred}{rgb}{0.7,0,0}

\usepackage{tikz}
\newcommand{\Cross}{$\mathbin{\tikz [x=1.4ex,y=1.4ex,line width=.3ex] \draw (0,0) -- (1,1) (0,1) -- (1,0);}$}
\newcommand{\Plus}{$\mathbin{\tikz [x=1.4ex,y=1.4ex,line width=.3ex] \draw (0,0) -- (1.4,0) (0.7,0.7) -- (0.7,-0.7);}$}
\newcommand{\OTimes}{$\mathbin{\tikz \draw[line width=.3ex]  (0,0)  circle (2.0mm);}$}

\newcommand\anote[1]{\textcolor{flamingo}{\bf [Ali:\,#1]}}
\newcommand\snote[1]{\textcolor{blue}{\bf [Shahin:\,#1]}}

\begin{document}

\begin{titlepage}

\begin{flushright}
{ IPM/P-2020/010\\
\today }\end{flushright}
\vspace{1cm}

\newcommand{\mytitle}{{Observable Quantum Loop Effects in the Sky}}


\begin{center}
\centerline{{\fontsize{20}{22} \selectfont {\textbf{ \mytitle}}}}
\vskip 2mm
%

\bigskip
\bigskip

{\bf{\large{A.A. Abolhasani}}}\footnote{e-mail:~\href{mailto:abolhasani@ipm.ir}{abolhasani@ipm.ir}}$^{; a,b}$, 
{\bf{\large{M.M.~Sheikh-Jabbari}}}\footnote{e-mail:~\href{mailto:jabbari@theory.ipm.ac.ir}{jabbari@theory.ipm.ac.ir}}$^{; b,c}$
\\

\normalsize
\bigskip

{$^a$ \it Department of Physics, Sharif University of Technology, Tehran, Iran}

{$^b$ \it School of Physics, Institute for Research in Fundamental Sciences (IPM)\\
P.O.Box 19395-5531, Tehran, Iran}


{$^c$ \it The Abdus Salam ICTP, Strada Costiera 11, 34151 Trieste, Italy}%



\end{center}
\setcounter{footnote}{0}

\medskip

\begin{abstract}

Expanding on \cite{Resonant-Monodromy}, we analyze in detail the single field chaotic inflationary models plus a cosine modulation term, augmented  by a light scalar field with inflaton dependent oscillatory mass term. We work out in detail the
Feynman diagrams and compute one, two and in general estimate higher loop two and three point functions in the in-in formulation. We explicitly establish how   the oscillatory mass term can amplify one-loop effects to dominate over the tree as well as the higher loop contributions.  The power spectrum of curvature perturbations of this model is hence enhanced compared to the simple single field chaotic model. As a consequence, one can suppress the tensor to scalar ratio $r$ and have a different expression for scalar spectral tilt and the running of the tilt, opening the way to reconcile chaotic models with convex potential and the  Planck data. As in monodromy inflation models, we also have a cosine modulation in the spectral tilt. We also analyze the bispectrum, which can be dominated by the amplified one-loop effects, yielding a new shape in non-Gaussianity. We discuss the bounds on parameter space from all available CMB observables and possible implications for reheating. 

\end{abstract}


\end{titlepage}

{\footnotesize
\tableofcontents
}
\section{Introduction}

Accumulating cosmological and in particular  CMB data \cite{Planck-2018, Akrami:2019izv} seems to match well with generic predictions of inflationary paradigm. There is a plethora of inflationary models in which various fields, typically scalar fields, are coupled to gravity, e.g. see \cite{Inflation-books}. The  most common class of inflationary models involve a scalar field with a potential which  is generically flat enough to derive slow-roll inflation. The simplest single-field inflationary models are chaotic models where the scalar is minimally coupled to gravity with a canonically normalized kinetic term and a polynomial potential. Despite the simplicity Planck data suggests that chaotic models with convex potentials are disfavored by the data \cite{Planck-2018}. 

Physical observables typically attributed to inflationary models are power spectrum of curvature perturbations and its spectral tilt and the bispectrum. These are related to two and three points of superhorizon inflaton perturbations. The general lore is that in a slow-roll inflationary model the loop contribution to these two and three point functions are generically small and it suffices to consider the tree level results \cite{Loop-Inflation-1}.

In our previous paper \cite{Resonant-Monodromy} we discussed a setup which defies the above lore. This model is an extension of the usual monodromy inflation model \cite{monodromy-SW,Chen:2008wn}, see also \cite{monodromy-MSWW, monodromy-and-Planck,Ngy-axion-monodromy, axion-monodromy-CMB-features, axion-monodromy-GW,Wenren:2014cga}, which has a cosine-modulated chaotic inflationary potential.  We couple this model to a ``amplifying module'', a light scalar field. In this model the one-loop two-point  function of scalar modes can be comparable or even slightly larger than the tree level result. 
In this work, we  elaborate further on this model and study and discuss in detail why and how this peculiar feature arises and whether the ``power spectrum amplification'' can jeopardise perturbative loop expansion. To carry out our computations we need to considerably extend and develop the quantum field theory techniques in the in-in formulation on an expanding (almost de Sitter) background \cite{Weinberg:2005vy}. In particular, we build upon the analysis of \cite{Weinberg:2005vy,Musso:2006pt} and further develop Feynman diagram technology in such models.

This work is organized as follows. In section \ref{sec:2}, we briefly review  our extended monodromy model, i.e. chaotic models with cosine modulation potential amended by a light scalar field coupled to it. In section \ref{sec:3}, we introduce basic Feynman rules of our model, setting the stage for loop analysis of the following sections. In section \ref{sec:4}, we present in detail the computation of one-loop two-point function of scalar perturbations. In section \ref{sec:5}, we present our one-loop three point function analysis. In section \ref{sec:6}, we discuss a generic higher loop analysis and establish validity of perturbative loop expansion. In section \ref{sec:observables},  we match our loop-corrected results for power spectra and bispectrum with the Planck data \cite{Planck-2018, Akrami:2019izv}. In section \ref{sec:discussion}, we conclude by summarizing and discussing our results. In some appendices we have gathered details of our loop computations. In appendices \ref{appen:two-pint-one-loop}, \ref{Appen:bispectrum-full} and \ref{appen:two-loop-2pt-funcn}, we show details of one-loop two and three point function, and  two-loop two-point function computations.
In appendix \ref{appen:loop-insertion}, we discuss construction of generic $L+1$-loop diagram out of given $L$-loop ones.

 \section{Extended monodromy and the amplification module}\label{sec:2}

The action for resonating extended monodromy inflation has two parts, an inflationary part and an amplification module. The inflationary part is a generalized axion monodromy inflation
in which the main part of the potential is  a generic power law of the inflaton field $\phi$ (rather than just the linear potential, as in original monodromy model \cite{monodromy-SW,axion-monodromy-GW}) with an added cosine modulations potential term. The amplification module is a light scalar field $\chi$ coupled to the inflaton field $\phi$. The full Lagrangian of the model is then
\be\label{full-action}
{\cal L}={\cal L}_\phi + {\cal L}_{\chi},
\ee
where the inflationary  and the amplification sectors, respectively ${\cal L}_\phi$ and ${\cal L}_\chi$,  are \cite{Resonant-Monodromy}
\be\label{L-phi}
{\cal L}_\phi=-\frac12(\partial\phi)^2-V_0(\phi),\qquad V_0(\phi)=\Lambda^4\left( (\frac{\phi}{ f})^p+b\cos(\frac{\phi}{f})\right),
\ee
\be\label{L-chi} 
{\cal L}_{\chi} = -\frac12(\partial\chi)^2- \frac12 m^2_{\chi}(\phi) \chi^2.
 \ee
In what follows we discuss each of these sectors in more detail.

\subsection{Inflationary background}\label{sec:2.1}

The inflationary potential  \eqref{L-phi} has two parameters of dimension of mass, $\Lambda, f$ and two dimensionless parameters $b$ and $p$. The $\phi^p$ part derives inflation and the coefficient of the oscillatory $\cos(\phi/f)$ is chosen such that 
it carries less than a percent of the potential energy of the inflaton, i.e. $b\lesssim 10^{-2}$. Therefore, it does not change the background inflationary trajectory in any essential way. It, however,  induces modulations on the observable signals like on the spectral tilt or non-Gaussianity which makes the model interesting and appealing.

The original monodromy model \cite{monodromy-SW,axion-monodromy-GW} has $p=1$ and has strong motivations from string and brane theory. Models with $p<1$ values are preferred by the current Planck data \cite{Planck-2018} and can also be motivated by the theoretical string theory considerations \cite{monodromy-and-Planck, monodromy-MSWW, axion-monodromy-CMB-features}. Nonetheless, one may not restrict oneself to such $p$ values. In fact, as we will argue amplified loop effects from the $\chi$ sector can reconcile $p>2$  convex potentials  with the Planck data. 

\paragraph{Background slow-roll dynamics.} The background evolution of inflaton in the slow-roll approximation is governed by 
\be\label{monodromy-slow-roll}
\dot\phi\simeq -\frac{p}{3}\frac{\Lambda^4}{f H}\left(\frac{\phi}{ f}\right)^{p-1},\quad\qquad 3M_{Pl}^2H^2\simeq\Lambda^4 \left(\frac{\phi}{f}\right)^p,
\ee
where  $\dot X$ denotes derivative w.r.t. comoving time $t$ and $H=\dot a/a$, $a(t)$ being the scale factor. Validity of the slow-roll approximation can be quantified through the slow-roll parameters
\be\label{slow-roll-parameters}
\begin{split}
\epsilon&=\frac12\left(\frac{M_{Pl}V_0'}{V_0}\right)^2\simeq \frac{p^2}{2}(\frac{M_{Pl}}{\phi})^2,\\ \eta&=\frac{M_{Pl}^2V_0''}{V_0}\simeq p(p-1)\left(\frac{M_{Pl}}{\phi}\right)^2- b \left(\frac{M_{Pl}}{\phi}\right)^p \left(\frac{M_{Pl}}{f}\right)^{2-p}\cos\frac{\phi}{f}.
\end{split}
\ee
The number of e-folds of the model is given by the usual equations
\be\label{Ne-epsilon}
N_e=\int_{\phi_f}^{\phi_i} \frac{1}{\sqrt{2\epsilon}} \frac{d\phi}{M_{Pl}},\qquad \phi\simeq \sqrt{2pN_e(\phi)} M_{Pl}=\frac{p}{\sqrt{2\epsilon}}M_{Pl}.
\ee
As in any chaotic (large field) model the inflaton $\phi$ has a super-Planckian roaming $\Delta\phi\sim 10M_{Pl}$. The energy scales $\Lambda, f$ are not fixed from the theoretical setting and should be determined upon the requirement of having a successful inflation model which yield $f,\Lambda\sim 10^{-3}-10^{-2}\ M_{Pl}$  \cite{axion-monodromy-GW, axion-monodromy-CMB-features}.  Therefore, during inflation $H\ll\omega\lesssim f\ll M_{Pl}$,

\paragraph{Approximate shift symmetry.} Single field slow-roll inflationary models generically enjoy an approximate continuous shift symmetry $\phi\to\phi+D$ for $D\lesssim M_{Pl}$. The characteristic of the periodic part of the inflaton potential $V_0(\phi)$ is that it is exactly invariant under discrete shift symmetries by integer multiples of $2\pi f$. The periodic part is the usual perturbatively protected axion potential term; this potential is induced through integrating out the corresponding instantons. As a result the continuous shift symmetry of the axion field is broken to a discrete shift-symmetry, $\phi\to\phi+2\pi f\mathbb{Z}$. 
As the inflaton roles down the $\phi^p$ potential, the oscillatory part oscillates with frequency $\omega=\dot\phi/f$. The parameter $\alpha$, 
\be\label{alpha-H/w}
\alpha\equiv \frac{\omega}{H}=\sqrt{2\epsilon}\frac{M_{Pl}}{f},
\ee
measures the number of times the inflaton oscillates  during inflation. During slow-roll evolution $\omega$, $H$ and hence $\alpha$ are almost constants 
and for the mentioned range of parameters,  $\alpha$ is of order $100$.

\paragraph{Observational features of monodromy inflation.} This model as usual large field models has a sizable tensor-scalar ratio $r$, $r\sim 0.007$, nonetheless, it does not suffer from super-Planckian field-roaming issue \cite{axion-monodromy-GW}. It has a controlable embedding in string theory for the same range of parameters which yields to successful inflation (see, however, \cite{Andriot-2015, Irene-2016}).

The oscillatory part of potential induces oscillatory patterns in the spectral tilt with the amplitude $\delta n_s\sim b \left(\frac{H}{\omega}\right)^{1/2}\ll 1$ as well as the resonant modulations of the cosine-log form  on the bispectrum and on $f_{NL}$ with amplitude $f_{res}\sim b \left(\frac{\omega}{H}\right)^{3/2}$ \cite{axion-monodromy-CMB-features, monodromy-and-Planck, Planck-2018}. These may be used to obtain some observationally viable range for $\alpha$.  A $\chi^2$ comparison with the minimal six parameter $\Lambda$CDM model has been performed, yielding  hints for such oscillatory templates for some frequencies in the range 
$\log_{10}(\alpha)\sim 1.5-2.1$ with $\Delta\chi^2\sim -10$ \cite{Resonant-Monodromy, Planck-2018, omega-bound}. 

\subsection{Amplification module}\label{sec:2.2}

Let us now discuss the modulating term ${\cal L}_\chi$. While the $\chi$ fields may have self interactions we consider them to be weakly coupled and ignore such interactions. The mass term for $\chi$, however, has a bare part $\mu$ which we take to be light  ($\mu \lesssim H$) and an induced part through interactions with the inflaton field $\phi$. Within a Wilsonian effective field theory description, the most general form of such an induced mass term which respects the Lorentz symmetry, the discrete shift symmetry of inflaton field $\phi$ and the $Z_2$ symmetry $\phi\to-\phi$ is,
\be\label{m-G}
m^2_{\chi}(\phi) = \mu^2+ f^2 \sum_{n=1} G_n((\partial_{\mu} \phi)^2/f^4) \cos^n \phi/f\,,
\ee
where $G_n(x)$ are polynomials in $x$. The above should be viewed as an effective potential valid for $(\partial_{\mu} \phi)^2/f^4\ll 1$ and hence to leading order
\be 
\label{chi:Mass}
m^2_{\chi}(\phi) = \mu^2+\mathcal{G} f^2 \cos \phi/f ,\qquad \mathcal{G}\equiv \sum_{n=1} g_n \left((\partial_{\mu} \phi)^2/f^4\right)^{n-1}.
\ee
where ${\cal G}$ is coefficient $G_1$ in \eqref{m-G}. In particular, for an inflationary background $(\partial_{\mu} \phi)^2=\dot\phi^2$ and $\gamma$,
\be\label{gamma-def}
\gamma\equiv \dfrac{\dot{\phi}}{f^2}=\frac{\omega}{f},
\ee
is almost a constant during the slow-roll period.  In general $\mathcal{G}=\mathcal{G}(\gamma)$ and we have a valid effective field theory description if $\gamma\lesssim 1$.

Modulated mass term for the $\chi$ field \eqref{chi:Mass} can affect both the background inflationary trajectory and the cosmic perturbation theory. In the next subsection we discuss the former. The effects on the cosmic perturbations are however very profound and will be discussed in the following sections.


\subsection{Particle production and stability of inflationary background}\label{sec:2-3}

The equation of motion for mode $\k$ of the $\chi$ field is
\be\label{chi-eom}
\ddot{\chi}_{_{\k}}+3 H\dot {\chi}_{_{\k}}+(k_{\mathrm{phys.}}^2+m_\chi^2(\phi)){\chi}_{_{\k}}=0,
\ee
where as usual $k_{\mathrm{phys.}}=k/a$. We take the $\chi$ field to be light in the sense that the oscillatory part in \eqref{chi:Mass} generically dominates over the $\mu^2$ part.\footnote{A similar case but with a non-derivative modulation has been studied in \cite{mehrdad}, see also \cite{Garcia:2020mwi,Amin:2017wvc}. As it becomes clear shortly, we are interested in studying relativistic modes while in \cite{mehrdad} heavy fields has been considered.} Due to the field-dependence of the mass, as the inflaton field $\phi$ rolls towards the minimum of the potential $V_0$, the matter field $\chi$ experiences a mass modulation with the same frequency as the background inflaton $\omega$. This time-dependent oscillatory mass term then yields  resonant $\chi$-particle production. 

After a rescaling of time variable $t$ to $z$, $z=\omega \,t/2$, \eqref{chi-eom} takes the form 
\be\label{eom-chi-II}
\chi ''_{_{\k}}+  \dfrac{6}{\alpha} \chi '_{_{\k}}+\left(A_{\k}+2 q \,\cos 2z \right)\chi_{_{\k}}=0,
\ee
where the primes are derivatives w.r.t. $z$, $\alpha$ is defined in \eqref{alpha-H/w} and
\be
A_{\k} :=4 \dfrac{k_{\mathrm{phys.}}^2}{\omega^2}+\frac{4\mu^2}{\omega^2}, \qquad q := \frac{2\mathcal{G}f^2}{\omega^2}.
\ee
For our case $q\ll 1$ and  $\alpha\gg 1$ and one can typically ignore the $\chi'$ term and \eqref{eom-chi-II} reduces to the canonical form of Mathieu equation,  
\be\label{Chi:Math}
\chi ''_{_{\k}}+\left(A_{\k}+2 q \,\cos 2z \right)\chi_{_{\k}}\simeq 0.
\ee

In general, solutions of this equation are classified into stability and instability bands according to the value of the so-called Floquet exponent $\mathcal{F}_k$. Floquet theorem states that the two solutions of the Mathieu equation is a product of $\exp(\pm\mathcal{F}_k z)$ and a periodic function, such that the product of two solutions is a periodic function in $2z$ \cite{Mathieu-Eq}. The Floquet index in principle is a complex number and when its real part is nonzero we have a exponentially growing (unstable) mode.  Particularly, in the language of stability-instability bands of Mathieu's differential equation, $q\ll 1$ corresponds to ``narrow resonance'' instability band \cite{Mathieu-Eq, preheating}.
For $q\ll 1$ resonant particle production occurs in narrow bands for 
\be
A_{\k}=l^2, \qquad l =1,2,3,\cdots
\ee
with width $\Delta A_{\k}\sim q^l$. So, the widest and most  growing solution occurs for $l=1$. Therefore, assuming $\mu\ll \omega$, we get
\be\begin{split}
k_{\mathrm{res.}} =& \dfrac{\omega}{2}+ \Delta k, 
\\
\dfrac{\Delta k}{k}=&\dfrac{1}{2} \dfrac{\Delta A_{\k}}{A_{\k}}= \dfrac{q}{2A_{\k}}= \dfrac{q}{2}.
\end{split}\ee
For the first instability band, the amplitude of the $\chi_{\k}$ is exponentially enhanced as \cite{Mathieu-Eq}
\be
\chi_{_{\k}} \sim e^{\mathcal{F}_{\k} z} \sim e^{q z/2}.
\ee
However, recalling expansion of the universe, every mode with comoving momentum $k$ soon leaves the narrow band in a time $\Delta t$,
\be
\label{Dt-Res}
\Delta t \sim \dfrac{2}{H} \dfrac{\Delta k}{k}= \dfrac{q}{H}.
\ee
Noting that $k_{\mathrm{phys.}} = k/a$,  modes get stretched so that they meet the resonance condition $A_\k =1 \pm q$ and subsequently grow exponentially. However, each mode soon exits the resonance when it is redshifted and $\Delta A_\k$ exceeds the resonance band width $q$. Therefore, there are always some modes which enter and some which leave  the narrow  resonance band. 

In the narrow resonance regime, the evolution of $\chi_{_{\k}}$ is seemingly adiabatic which gives rise to slow growth of the amplitude of the oscillations of $\chi$ field. Having found the solutions to the $\chi$ equation of motion, we can quantize the field using canonical quantization by imposing Bunch-Davies vacuum initial condition ($\alpha_{\K}=1$ and $\beta_{\K}=0$ in the equation below). To our approximation, $\chi_{_\K}$ for the time interval of the resonance is
\be 
\label{Math-sol} 
\chi_{_{\k}}  \simeq \dfrac{1}{\sqrt{2k}}\Big[\alpha_{\K}(z) \,e^{-i \sqrt{A_{\K}} z} + \beta_{\K} (z) \, e^{+i \sqrt{A_{\K}} z} ~\Big] 
\ee
in which
\be
\alpha_{\K} (z) = \cosh\,\mathcal{F}_{\K} z,\qquad \beta_{\K}(z)= \sinh \,\mathcal{F}_{\K} z.
\ee
For the first instability band $A_\k=1$ and the Floquet index is found to be 
\be
 \exp(\mathcal{F}_{\K}z) \simeq \exp \left[ \left(\dfrac{q}{2} -\dfrac{\Delta k^2}{k_{\mathrm{res.}}^2} \right) \dfrac{\omega t}{2} \right].
\ee
As a result, number density of resonantly produced $\chi$ particles $n_{\chi}$ is\footnote{Note that the relation employed in \cite{preheating},  $n_{\K}=\dfrac{\omega_{\K}}{2} \left( \dfrac{\vert\dot{\delta \chi}_{_{\k}}\vert ^2}{\omega_{\K}^2}+\vert \delta \chi_{_{\k}}\vert ^2 \right)$, 
estimates the number of particles twice as our rigorous result \eqref{n-K}.}
\begin{align}
\label{n-K}
 n_{\K} = \vert \beta_{\K}\vert^2 = \sinh^2 (\mathcal{F}_\k \frac{\omega \Delta t}{2})=\sinh^2 (\frac{q^2\alpha }{4}),
\end{align}

\paragraph{Back-reaction on background inflationary trajectory.} While there is a burst of $\chi$ particle production in the narrow band resonance,  the exponential expansion of the universe dilutes the $\chi$  particle.  The energy density of $\chi$-particles 
is given by
\begin{align}
\label{chi-energy:1}
\nonumber
\rho_{\chi} = \dfrac{1}{2\pi^2 a^3} \int_{0}^{a(t) \omega/2}   dk \,k^2 \, \omega_\k \, n_\k &= \dfrac{
\sinh^2(\frac{q^2\alpha}{4})}{2\pi^2 a^3} \int_{0}^{a(t) \omega/2}  \md k\, k^2 \, \dfrac{k}{a} 
\\
&=  \dfrac{\omega^4}{128 \pi^2} \, \sinh^2(\frac{q^2\alpha}{4}).
\end{align}
That is, energy density transferred to highly relativistic $\chi$ particles (radiation)  reaches a constant value during inflation. 
For the range of parameters we are interested in $q^2\alpha\ll 1$ and hence 
\be
\rho_{\chi} \simeq \dfrac{\omega^4}{128 \pi^2} \, (\frac{q^2\alpha}{4})^2= \frac{q^4\alpha^6}{512 \pi^2} H^4.
\ee
The effects of the particle production on the background inflationary trajectory is governed by the ratio
\be
\label{energy-ratio:1}
R\equiv\dfrac{\rho_{\chi}}{\rho_{\chi}+\rho_{\phi}} = \frac{\frac{q^4\alpha^6}{512 \pi^2} H^2}{3M_{Pl}^2+\frac{q^4\alpha^6}{512 \pi^2} H^2}. 
\ee
For our typical range of parameters $q\sim 10^{-2}$ and $\alpha^2\sim 10^3-10^4$, therefore, $R\lesssim 10^{-1}(H/M_{Pl})^2\sim 10^{-10}\lll 1$.
So, we can safely ignore the back-reaction effects during inflation. While not relevant during inflation, the particle production could become important towards the end of inflation to provide us with a (p)reheating setting. We shall comment on the latter in the last section.

\section{Modulated  interactions and cosmic perturbations}\label{sec:3}

In the previous section we introduced the generalized monodromy inflation model plus a light scalar with modulated mass term and discussed the classical inflationary background trajectories. In this section we set the basics of  cosmic perturbation theory of this model and introduce the basic Feynman graphs for the field theory computation in the in-in formulation.

\subsection{Modulated  interactions}

As discussed the classical background trajectory of the $\phi$ and $\chi$ fields are
\be\label{classical-path}
\phi\simeq \phi_i+f\omega (t-t_0),\qquad \chi\simeq 0,
\ee
where $\simeq$ means to leading slow-roll order. We choose the initial time $\omega t_0=\phi_i/f$. We analyze pertubations around the classical path \eqref{classical-path} and denote quantum subhorizon  perturbations of inflaton by $\varphi$, $\phi=f\omega t+\varphi$ and those of the modulating field by $\chi$. The action for the scalar perturbations in the  spatially flat gauge takes the form,
\be\label{L-phi-chi-pert}
{\cal L}=-\frac12(\partial\varphi)^2-\frac32\eta H^2 \varphi^2-\frac12(\partial\chi)^2-\dfrac{1}{2} {\cal G} f^2 \cos (\omega t) \chi^2+{\cal L}_{\mathrm{int.}}+ {\cal O}(\varphi^3),
 \ee
where we used \eqref{slow-roll-parameters} to replace for $V''_0$ in terms of slow-roll parameter $\eta$ and   
\be\label{interaction-Hamiltonian}
{\cal L}_{int.} \simeq -g_3(t) \vphi \chi^2- g_4(t) \vphi^2\chi^2,
\ee
with
\be\label{g3-g4}
g_3(t)=-\frac12{\cal G} f  \sin \omega t,\qquad g_4(t)=-\frac14 {\cal G}  \cos \omega t.
\ee

As in any perturbative quantum field theory we study \eqref{L-phi-chi-pert} pertubatively in powers of ${\cal G}$. The first step toward this is to consider the free theory. The equation of motion for $\chi$ is \eqref{chi-eom} and for $\varphi$ is, 
\be\label{varphi-eom}
\ddot\varphi+3H\dot\varphi-\nabla^2\varphi+ 3H^2\eta \varphi=0.
\ee
It is more convenient to analyze these equations in the conformal time
\be\label{conformal-time}
\tau=-\frac{e^{Ht}}{H},
\ee
where the background metric takes the form
\be
ds^2=\frac{1}{H^2\tau^2} (-d\tau^2+d\y\cdot d\y).
\ee
Note that the conformal time ranges in $\tau \in(-\infty, 0)$. In particular it is always negative and  $H\tau\to 0$ ($H\tau \to -\infty$) covers superhorizon (deep inside horizon) regions.

One may Fourier-expand $\varphi$ in terms of Fourier modes $\varphi(\k,\tau)$. Solution to \eqref{varphi-eom}, to the leading order in slow-roll parameter is
\be\label{varphi-mode}
\varphi(\k,\tau) = \dfrac{H}{\sqrt{2 k^3}} (i-k\tau) e^{-ik\tau},
\ee
where the initial condition is given by the Bunch-Davies vacuum. That is, at early times ($|k\tau|\gg 1$) when the modes are deep inside horizon, it simplifies to 
\be\label{DIH-varphi}
\varphi(\k,\tau) \rightarrow \dfrac{-H \tau }{\sqrt{2 k}} e^{-ik\tau}.
\ee

In section \ref{sec:2-3} we have analyzed the dynamics of  $\chi$ particles ignoring the interactions with $\varphi$ particles. The result for the mode function we found, in the Bunch-Davies vacuum state is
\be\label{chi-mode-expansion-1}
\chi(\tau,x)=\int \frac{d^3p}{(2\pi)^3} \ \left(a(\tau) \chi_{\p}(\tau) e^{i\p\cdot x}+ c.c.\right)
\ee
where
\be\label{chi-mode-expansion-2}
a(\tau)\chi_{\p}(\tau) =
    \begin{cases}
    \frac{1}{\sqrt{2\omega_p(\tau)}} \,\left(\alpha_p \, f_{+}(\p,\tau)+ \beta_{p}\,f_{-}(\p ,\tau)\right),&   p\tau \leq \frac{\alpha}{2}  \\
    \frac{1}{\sqrt{2\omega_p(\tau)}}\,  f_{+}(\p,\tau)&  \mathrm{otherwise}
    \end{cases}
 \ee
where in our notation $p^2=\p\cdot \p=|\p|^2$ and the adiabatic mode functions $f_{\pm}$ are
\be
f_{+}(\p,\tau)=f_{-}^{\ast}(\p,\tau) = \exp \left( -i \int^{\tau} \omega_p(\tau') \,d\tau' \right)
\ee
in which $\omega_p(\tau) = \sqrt{p^2+\frac{\mu^2}{H^2\tau^2}}\simeq p$.  As discussed in section \ref{sec:2-3}, $\beta_p$ coefficients show the particle production due to having time-dependent mass term. The particle production is happening in the narrow-band resonance and is not very significant effect during inflation. Therefore, in our quantum field theory analysis below we ignore the $\beta_p$ term in the $\chi$-modes and in the end discuss how good this approximation is.



\subsection{In-in perturbation theory and Feynman rules}\label{sec:3-2}

To make a systematic analysis of one, two and higher loops of two and three point functions, in this part we layout the basic tools. These analyses may be facilitated by the Feynman diagram notation that we also develop. Since we are using the in-in formulation, the Feynman diagram notion is more involved than the  more familiar in-out formulation. To draw the Feynman diagrams we use the following conventions as in  \cite{Musso:2006pt}:
\begin{itemize}
    \item Bullets carry the coordinate at which fields are calculated.
    \item Arrows show the direction of increase of conformal time $\tau$. Note that with \eqref{conformal-time}, $\tau\leq 0$.
    \item  Black and red {solid lines} respectively correspond to the propagator of $\vphi$ and $\chi$. 
    \item {Dashed line} with a cross at one end denotes a free field evaluated at the argument of the other end without the cross. 
\end{itemize} 
\paragraph{Basic Feynman diagrams.} 
 \begin{itemize}\item  Fields at zeroth order, $\vphi_0, \chi_0$:
  \begin{figure}[H]
    \begin{axopicture}(170,30)(-100,-10)
    \SetWidth{1.5}
    \SetColor{Black}
    \DashLine(40,0)(0,0){2}
    \Text(40,10){$\y$}
     \Text(0,0){\Cross}
    \Text(65,0){$= \vphi_0(\y)$}
\end{axopicture}
\begin{axopicture}(170,30)(-170,-10)
    \SetWidth{1.5}
    \SetColor{Red}
    \DashLine(40,0)(0,0){2}
     \Text(0,0){\Cross}
    \Text(40,10){$y$}
    \SetColor{Black}
    \Text(65,0){$= \chi_0(y)$}
\end{axopicture}
    \end{figure}

\item  The propagator $G_{\vphi}(\tau,\tau';\x,\x')$ is retarded Green's function of $\vphi$ field, 
\be\label{propagator-phi}
\begin{split}
G_{\vphi}(\tau,\tau';\x,\x')&=i\theta(\tau-\tau') \big[\vphi(\tau,\x),\vphi(\tau',\x')\big]\\
&= iH^2\int \frac{d^3\k}{2k^3} \ \theta(\tau-\tau')e^{i\k\cdot (\x-\x')} \big[e^{-ik(\tau-\tau')}(i-k\tau)(i+k\tau')- c.c. \ \big].
\end{split}
\ee
Similarly, the $\chi$-field propagator 
$G_{\chi}$ is the retarded Green's function of $\chi$ perturbations
\be\label{chi-propagator}
G_{\chi}(\tau,\tau';\x,\x')=i\theta(\tau-\tau') \big[\chi(\tau,\x),\chi(\tau',\x')\big].
\ee
Recalling the mode expansion of $\chi$, \eqref{chi-mode-expansion-1} and \eqref{chi-mode-expansion-2}, $G_\chi$ in \eqref{chi-propagator} has a term as in \eqref{propagator-phi} and terms proportional to particle production coefficients $\beta_k$ and $\beta_k^2$. As discussed in the previous section, however, during inflation the particle production, which happens in the narrow band resonance, is not significant and  one may hence safely approximate $G_\chi$ with the same expression as in \eqref{propagator-phi}. Pictorially they may be depicted as
    \begin{figure}[H]
    \begin{axopicture}(170,40)(-100,-10)
    \SetWidth{1.5}
    \SetColor{Black}
    \Text(0,10){$\x$}
    \Text(40,10){$\y$}
    \ArrowLine(40,0)(0,0)
    \Text(75,0){$=G_{\vphi}(\x,\y)$}
\end{axopicture}
\begin{axopicture}(170,40)(-170,-10)
    \SetWidth{1.5}
    \SetColor{Red}
    \ArrowLine(40,0)(0,0)
    \SetColor{Black}
    \Text(0,10){$\x$}
    \Text(40,10){$\y$}
    \Text(75,0){$=G_{\chi}(\x,\y)$}
\end{axopicture}
    \end{figure}
{Note that both $\varphi$ and $\chi$ fields are light, essentially massless fields in our analysis, and hence have similar propagators.}
\item We have cubic and quartic interaction terms \eqref{interaction-Hamiltonian}. The quartic $\chi^2\varphi^2$ term, however, may be neglected in our analysis, as it will not lead resonances which enhance contributions of the cubic interactions. So,  we only focus on the cubic interaction term. Each vertex at $y$ corresponds to a spacetime integral, which may have solid or dashed lines attached to it. Here we show one with solid lines but either of these lines can be solid or dashed.
\begin{figure}[H]
    \begin{axopicture}(170,80)(-170,-30)
    \SetWidth{1.5}
    \SetColor{Black}
    \Line(40,0)(0,0)
     \SetColor{Red}
     \Line(40,0)(68,28)
    \Line(40,0)(68,-28)
    \Text(40,10){$\y$}
    \SetColor{Black}
    \Text(178,0){$= \displaystyle -\int d^4\y g_3(\tau_y)=-\int d^3\y\ d{\cal T}_y$}
\end{axopicture}
    \end{figure}
\end{itemize}
where we have defined
\begin{equation}\label{d-cal-T}
    d{\cal T} \equiv \dfrac{d \tau}{H^4 \tau^4}\,g_3(\tau).
\end{equation}

\paragraph{Generic cosmological equal time $m$-point operators, perturbative treatment.}

In the cosmological perturbation theory, in general, we are interested in the expectation value of products of various fields, all {at the same time} but generally with different spatial arguments \cite{Weinberg:2005vy}. In particular, we want to calculate the {equal time} correlation function of cosmological observables, e.g. curvature perturbations. It is well known that product of $m$ field operators, e.g. $\hat{O}(\tau)= \vphi(\tau;\x_1) \vphi(\tau;\x_2) ... \vphi(\tau;\x_m)$, to the $V$-order in perturbation theory can be expanded as\footnote{Note that $V$ order in perturbation theory corresponds to a Feynman graph with $V$ number of vertices.}
\be
\label{O-N:exp}
\hat{O}(\tau) = \sum_{V} \hat{O}_V(\tau), \qquad \hat{O}_V= \sum_{i_1+\cdots+i_m=V} \vphi_{i_1}(\tau;\x_1) \vphi_{i_2}(\tau;\x_2) \cdots \vphi_{i_m}(\tau;\x_m)
\ee
where $\vphi_{i}$ denotes the $i$-the order fluctuation of the operator $\vphi$. In  particular, for equal time product of two scalar field $\hat{O}(\tau)= \vphi(\tau;\x_1) \vphi(\tau;\x_2)$ we get 
\be
\label{two-point:N}
\hat{O}_{V}= \sum_{i=0}^{V} \vphi_{i}(\tau;\x_1)\,\vphi_{V-i}(\tau;\x_2)
\ee
So, we need to know $\vphi$ at $i$-order in perturbation theory. To this end, we recall that \cite{Weinberg:2005vy,Musso:2006pt}
\be
\label{pert-nested:eq1}
\hat{O}_V(\tau) = i^V \int \prod_{i=1}^{V} \dfrac{d \tau_i}{H^4 \tau_i^4} \theta(\tau_{i-1}-\tau_i) \int \prod_{i=1}^V d^3 \y_i \bigg[{\cal H}_{int}(\tau_V,\y_V),...,\bigg[{\cal H}_{int}(\tau_2,\y_2),\bigg[{\cal H}_{int}(\tau_1,\y_1),\hat{O}_0(\tau)\bigg]\bigg]
\ee
where ${\cal H}_{int}=-{\cal L}_{int}$ denotes the interaction Hamiltonian operator, in our case given in \eqref{interaction-Hamiltonian}.

In  one-loop two and three point function analysis, we need to know
$\varphi_1, \varphi_2, \vphi_3$. Eq.\eqref{pert-nested:eq1} can be used to find $\vphi_1$ 
 \begin{figure}[h!]
 \begin{axopicture}(170,70)(-150,-30)
    \SetWidth{1.5}
    \SetColor{Black}
    \Text(0,10){$\x$}
    \Vertex(0,0){3}
    \ArrowLine(50,0)(0,0)
    \Text(43,8){$\y_1$}
    \SetColor{Red}
    \DashCArc(80,0)(30,90,270){3}
    \SetColor{Red}   
    \Text(80,30){\Cross}
    \Text(80,-30){\Cross}
   \SetColor{Black}
    \Text(-33,0){$\vphi_1(\tau,\x) =$}
\end{axopicture}
\end{figure}
\be\label{varphi-1}
\hspace{-1.4cm} =- \int d^3\y_1\, d{\cal T}_1\ G_{\vphi}(\tau,\tau_1;\x,\y_1) \chi^2 (\tau_1,\y_1)
\ee
where 
$G_\vphi$ is given in \eqref{propagator-phi}. Similarly $\vphi_2$ is
\begin{figure}[H]
\begin{axopicture}(170,80)(-100,-35)
    \SetWidth{1.5}
    \SetColor{Black}
    \Vertex(0,0){3}
    \ArrowLine(50,0)(0,0)
    \SetColor{Red}
    \ArrowArc(90,0)(40,90,180)
    \DashCArc(90,0)(40,180,270){3}
    \Text(90,-40){\Cross}
    \SetColor{Black}
    \DashCArc(110,40)(20,90,180){2}
    \SetColor{Red}
    \DashCArc(110,40)(20,180,270){2}
    \Text(110,20){\Cross}
    \SetColor{Black}
    \Text(110,60){\Cross}
    \Text(0,10){$\x$}
    \Text(44,9){$\y_1$}
    \Text(84,48){$\y_2$}
    \Text(-35,0){$\vphi_2(\tau,\x)  =$}
    \Text(140,0){$+$}
\end{axopicture}
\begin{axopicture}(170,120)(-110,-35)
    \SetWidth{1.5}
    \SetColor{Black}
    \Vertex(0,0){3}
    \ArrowLine(50,0)(0,0)
    \SetColor{Red}
    \DashCArc(90,0)(40,90,180){3}
    \ArrowArcn(90,0)(40,270,180)
    \Text(90,40){\Cross}
    \SetColor{Black}
    \DashCArc(110,-40)(20,90,180){2}
    \SetColor{Red}
    \DashCArc(110,-40)(20,180,270){2}
    \Text(110,-60){\Cross}
    \SetColor{Black}
    \Text(110,-20){\Cross}
    \Text(0,10){$\x$}
    \Text(44,9){$\y_1$}
    \Text(82,-50){$\y_2$}
\end{axopicture}
\end{figure}
\begin{align}\label{varphi-2}
\hspace{1.2cm}= 2\int d^3\y_1 d^3\y_2 \, d{\cal T}_1 d{\cal T}_2\ \vphi (\tau_2,\y_2)
G_{\vphi}(\tau,\tau_1;\x,\y_1) G_{\chi}(\tau_1,\tau_2;\y_1,\y_2) \{\chi (\tau_1,\y_1), \chi (\tau_2,\y_2)\},
\end{align}
in which $\{,\}$ denotes symmetrization of field operators (it is anticommutator). 
Finally, after simple but lengthy  algebra we find $\varphi_3$ 
\begin{figure}[H]
\begin{axopicture}(170,120)(-40,-35)
    \SetWidth{1.5}
    \SetColor{Black}
    \Vertex(0,0){3}
    \ArrowLine(50,0)(0,0)
    \SetColor{Red}
    \ArrowArc(90,0)(40,90,180)
    \DashCArc(90,0)(40,180,270){3}
    \Text(90,-40){\Cross}
    \SetColor{Red}
    \DashCArc(110,40)(20,90,180){2}
    \SetColor{Black}
    \ArrowArcn(110,40)(20,270,180)
    \SetColor{Red}
    \Text(110,60){\Cross}
    \SetColor{Red}
    \DashCArc(130,20)(20,90,180){2}
    \SetColor{Red}
    \DashCArc(130,20)(20,180,270){2}
    \Text(130,0){\Cross}
    \SetColor{Red}
    \Text(130,40){\Cross}
    \SetColor{Black}
    \Text(0,10){$x$}
    \Text(44,9){$y_1$}
    \Text(84,48){$y_2$}
    \Text(120,18){$y_3$}
    \Text(-30,0){$\vphi_3(\tau,\x) =$}
    \Text(150,0){$+$}
\end{axopicture}
\begin{axopicture}(170,80)(-40,-35)
    \SetWidth{1.5}
    \SetColor{Black}
    \Vertex(0,0){3}
    \ArrowLine(50,0)(0,0)
    \SetColor{Red}
    \ArrowArc(90,0)(40,90,180)
    \DashCArc(90,0)(40,180,270){3}
    \Text(90,-40){\Cross}
    \SetColor{Black}
    \DashCArc(110,40)(20,90,180){2}
    \SetColor{Red}
    \ArrowArcn(110,40)(20,270,180)
    \SetColor{Black}
    \Text(110,60){\Cross}
    \SetColor{Black}
    \DashCArc(130,20)(20,90,180){2}
    \SetColor{Red}
    \DashCArc(130,20)(20,180,270){2}
    \Text(130,0){\Cross}
    \SetColor{Black}
    \Text(130,40){\Cross}
    \SetColor{Black}
    \Text(0,10){$x$}
    \Text(44,9){$y_1$}
    \Text(84,48){$y_2$}
    \Text(120,18){$y_3$}
    \Text(150,0){$+$}
\end{axopicture}
\begin{axopicture}(170,120)(-50,-35)
    \SetWidth{1.5}
    \SetColor{Black}
    \Vertex(0,0){3}
    \ArrowLine(50,0)(0,0)
    \SetColor{Red}
    \ArrowArc(90,0)(40,90,180)
    \ArrowArcn(90,0)(40,270,180)
    \SetColor{Black}
    \DashCArc(110,-40)(20,90,180){2}
    \SetColor{Red}
    \DashCArc(110,-40)(20,180,270){2}
    \Text(110,-60){\Cross}
    \SetColor{Black}
    \Text(110,-20){\Cross}
    \SetColor{Black}
    \DashCArc(110,40)(20,90,180){2}
    \SetColor{Red}
    \DashCArc(110,40)(20,180,270){2}
    \Text(110,20){\Cross}
    \SetColor{Black}
    \Text(110,60){\Cross}
    \Text(0,10){$x$}
    \Text(44,9){$y_1$}
    \Text(82,-50){$y_3$}
    \Text(82,50){$y_2$}
\end{axopicture}
\end{figure}
\begin{align}
 = &\int d^3\y_1 d^3\y_2 d^3\y_3 \, d{\cal T}_1 d{\cal T}_2 d{\cal T}_3\ \nonumber \\ &\Big [
-2G_{\vphi}(\tau,\tau_1;\x,\y_1) G_{\chi}(\tau_1,\tau_2;\y_1,\y_2) G_{\vphi}(\tau_2,\tau_3;\y_2,\y_3) \big\{\chi (\tau_1,\y_1), \chi (\tau_2,\y_2) \big \} \chi^2 (\tau_3,\y_3)\nonumber
\\
\nonumber
&-4G_{\vphi}(\tau,\tau_1;\x,\y_1) G_{\chi}(\tau_1,\tau_2;\y_1,\y_2) G_{\chi}(\tau_2,\tau_3;\y_2,\y_3) \bigg\{\chi (\tau_1,\y_1), \big\{ \vphi (\tau_2,\y_2),\chi (\tau_3,\y_3) \vphi (\tau_3,\y_3) \big\} \bigg\}  
\\
&-4G_{\vphi}(\tau,\tau_1;\x,\y_1) G_{\chi}(\tau_1,\tau_2;\y_1,\y_2) G_{\chi}(\tau_1,\tau_3;\y_1,\y_3) \bigg \{\chi (\tau_3,\y_3) \vphi (\tau_3,\y_3), \chi (\tau_2,\y_2) \vphi (\tau_2,\y_2) \bigg \}  
 \,\Big ]
\label{varphi-3}
\end{align}




\section{{Two-point function, one-loop analysis}}\label{sec:4}

In this section we compute the two-point function $\expect{\vphi (\k,\tau) \vphi(\k',\tau)}$ to one-loop order. In the {spatially flat gauge} we have adopted here, this two-point function once computed for superhorizon modes, i.e. for $k \tau\to 0$, is proportional to the power spectrum of scalar perturbations. The results and analysis of this section and the next section paves the way to perform an all loop analysis, to which we return in section \ref{sec:6}.

\begin{figure}[ht]
\begin{axopicture}(70,220)(-180,-180)
    \SetWidth{1.5}
    \SetColor{Black}
    \Vertex(0,0){3.5}
    \ArrowLine(45,0)(0,0)
    \SetColor{Red}
    \DashCArc(82.5,0)(37.5,0,90){3}
    \SetColor{Red}
    \DashCArc(82.5,0)(37.5,90,180){3}
    \SetColor{Red}
    \DashCArc(82.5,0)(37.5,180,360){3}
    \SetColor{Red}
    \SetColor{Black}
    \ArrowLine(120,0)(165,0)
    \Vertex(165,0){3.5}
    \Text(80,-60){(a)}
    \Text(0,10){$\x_1$}
    \Text(160,10){$\x_2$}
    \Text(40,-9){$\tilde{\y}_1$}
    \Text(127,-9){$\tilde{\y}_2$}
    
    \SetColor{Red}
    \Text(82.5,-37.5){\Cross}
    \Text(82.5,37.5){\Cross}
  \end{axopicture} 
\begin{axopicture}(70,220)(30,-80)
    \SetWidth{1.5}
    \SetColor{Black}
    \Vertex(0,0){3.5}
    \ArrowLine(45,0)(0,0)
    \SetColor{Red}
    \ArrowArc(82.5,0)(37.5,0,180)
    \SetColor{Red}
    \DashCArc(82.5,0)(37.5,180,360){3}
    \SetColor{Black}
    \DashLine(120,0)(165,0){3}
    \Vertex(165,0){3.5}
    \Text(80,-60){(b)}
   \Text(142.5,0){\Cross}
   \SetColor{Red}
   \Text(82.5,-37.5){\Cross}
   \SetColor{Black}
   \Text(0,10){$\x_1$}
    \Text(160,10){$\x_2$}
    \Text(40,-9){$\tilde{\y}_1$}
    \Text(127,-9){$\tilde{\y}_2$}
  \end{axopicture} 
  \begin{axopicture}(70,220)(-180,-80)
    \SetWidth{1.5}
    \SetColor{Black}
    \Vertex(0,0){3.5}
    \DashLine(45,0)(0,0){3}
    \SetColor{Red}
    \ArrowArcn(82.5,0)(37.5,180,0)
    \SetColor{Red}
    \DashCArc(82.5,0)(37.5,180,360){3}
    \SetColor{Black}
    \ArrowLine(120,0)(165,0)
    \Vertex(165,0){3.5}
    \Text(80,-60){(c)}
   \Text(22.5,0){\Cross}
   \SetColor{Red}
   \Text(82.5,-37.5){\Cross}
   \SetColor{Black}
   \Text(0,10){$\x_1$}
    \Text(160,10){$\x_2$}
    \Text(40,-9){$\tilde{\y}_2$}
    \Text(127,-9){$\tilde{\y}_1$}
  \end{axopicture} 
\caption{One-loop contribution to the scalar power $\expect{\vphi(\k) \vphi(-\k)}$. (a) Expectation value of two first order perturbations $\expect{\vphi_1(\k) \vphi_1(-\k)}$ and (b), (c) that of a vacuum fluctuation and a second order fluctuation  $\expect{\vphi_0(\k) \vphi_2(-\k)}$ and $\expect{\vphi_2(\k) \vphi_0(-\k)}$.}\label{Power-spectrum:Feyn-Dia--One-loop}
   \end{figure}
\subsection{Contributing diagrams}\label{sec:4-1}
As depicted in Fig. \ref{Power-spectrum:Feyn-Dia--One-loop}, there are three terms contributing to the scalar power $P=\expect{\vphi \vphi}$  at second order,
\be
P_{\rm{1-loop}}= P_{11}+P_{20}+P_{02}=\expect{\vphi_1 \vphi_1} + \expect{\vphi_2 \vphi_0}+ \expect{\vphi_0 \vphi_2},
\ee
where $P_{11}, P_{20}, P_{02}$ are the superhorizon values ($k\eta\to 0$) of the Fourier mode of the following expressions:
\begin{align}
\nonumber
P_{11}(\tau,\x_1;& \tau,\x_2) \equiv \expect{\vphi_1(\tau,\x_1) \vphi_1(\tau,\x_2)}=
\\
&2\times \int d^3\tilde{\y}_1 d^3\tilde{\y}_2\ d{\tilde{\cal T}}_1d{\tilde{\cal T}}_2\ 
G_{\vphi}(\tau,\tilde{\tau}_1;\x_1,\tilde{\y}_1) ~ G_{\vphi}(\tau,\tilde{\tau}_2;\x_2,\tilde{\y}_2) \langle \hat{\chi} (\tilde{\tau}_1,\tilde{\y}_1) \hat{\chi} (\tilde{\tau}_2,\tilde{\y}_2) \rangle^2
\end{align}
in which the overall factor of $2$ is number of possible ways to contract internal dashed lines of the corresponding diagram, i.e. diagram (a) in Fig.\ref{Power-spectrum:Feyn-Dia--One-loop}.  The contribution of the diagrams (b) and (c)  are
\begin{align}
\nonumber
P_{20}(\tau,\x_1&;\tau,\x_2) \equiv \expect{\vphi_2(\tau,\x_1) \vphi_0(\tau,\x_2)}= 2\times
\int d^3\tilde{\y}_1 d^3\tilde{\y}_2\ d{\tilde{\cal T}}_1d{\tilde{\cal T}}_2\ 
\\
~&\,G_{\vphi}(\tau,\tilde{\tau}_1;\x_1,\tilde{\y}_1) G_{\chi}(\tilde{\tau}_1,\tilde{\tau}_2;\tilde{\y}_1,\tilde{\y}_2)  \,\big\langle \{\hat{\chi} (\tilde{\tau}_1,\tilde{\y}_1), \hat{\chi} (\tilde{\tau}_2,\tilde{\y}_2)\} \big\rangle \big \langle \hat{\vphi} (\tilde{\tau}_2,\tilde{\y}_2) \hat{\vphi}(\tau,\x_2) \big \rangle,
\end{align}
\begin{align}
\nonumber
P_{02}(\tau,\x_1&;\tau,\x_2) \equiv \expect{\vphi_0(\tau,\x_1) \vphi_2(\tau,\x_2)}= 2\times
\int d^3\tilde{\y}_1 d^3\tilde{\y}_2\ d{\tilde{\cal T}}_1d{\tilde{\cal T}}_2\ 
\\
~& G_{\vphi}(\tau,\tilde{\tau}_1;\x_2,\tilde{\y}_1) G_{\chi}(\tilde{\tau}_1,\tilde{\tau}_2;\tilde{\y}_1,\tilde{\y}_2)  \,\big\langle \{ \hat{\chi} (\tilde{\tau}_1,\tilde{\y}_1), \hat{\chi} (\tilde{\tau}_2,\tilde{\y}_2) \} \big\rangle \big \langle \hat{\vphi}(\tau,\x_1) \hat{\vphi} (\tilde{\tau}_2,\tilde{\y}_2)  \big \rangle.
\end{align}
In these expression we have defined 
\be
\hat{\chi}_{\p}(\tau)= \chi_{\p}(\tau)\boldsymbol{a}^\chi_{\p}+\chi^{\ast}_{-\p}(\tau)(\boldsymbol{a}^\chi)^{\dagger}_{-\p}\,,\qquad \hat{\vphi}_{\p}(\tau)= \vphi_{\p}(\tau)\boldsymbol{a}^\vphi_{\p}+\vphi^{\ast}_{-\p}(\tau)(\boldsymbol{a}^\vphi)^{\dagger}_{-\p}\,
\ee
where $\boldsymbol{a}_\p, \boldsymbol{a}^\dagger_\p$ are the usual creation, annihilation operators.

In particular, using above result, one may compute the  late time $k\eta\to 0$ behavior of the contributions to two-point functions:
\begin{align}
\nonumber
P_{11}(k)\equiv \expect{\vphi^{(1)}_{\k} \vphi^{(1)}_{-\k}}&= 8 P_{\phi}(k)\times\int d^3\p_1 d^3\p_2
\int d{\cal T}_1\ d{\cal T}_2\times\\
&\times\mathrm{Re} \vphi_{\k}(\tau_1) ~ \mathrm{Re} \vphi_{-\k}(\tau_2) \expect{\hat{\chi}_{\p_1}(\tau_1)\hat{\chi}_{-\p_2}(\tau_2)} \expect{ \hat{\chi}_{-\p_1+\k}(\tau_1) \hat\chi_{\p_2-\k}(\tau_2) },
\end{align}
where $d{\cal T}$ is defined in \eqref{d-cal-T} and 
\be\label{G-P-prop}
G_{\vphi}(0,\tau_1;\k) =-2 P_{\phi}^{1/2}(k)~\mathrm{Re} \vphi_{\k}(\tau_1),\qquad P_{\phi}(k):=\expect{\vphi^{(0)}_{\k} \vphi^{(0)}_{-\k}}_{k\eta\to 0}=\frac{H^2}{2k^3}.
\ee
Noting that $\expect{\hat{\chi}_{\p_1}(\tau_1)\hat{\chi}_{-\p_2}(\tau_2)}=\chi_{p_1}(\tau_1) \chi^{\ast}_{p_2}(\tau_2) \,\delta(\p_1-\p_2)$, one gets
\begin{align}\label{P11}
P_{11}(k)
=8 P_{\phi}(k)\int d^3\p \int d{\cal T}_1\ \mathrm{Re} \vphi_{k}(\tau_1) \int^{\tau_1} d{\cal T}_2\ \mathrm{Re} \vphi_{k}(\tau_2)\bigg[\Xi(\tau_1,\tau_2)+\Xi^{\ast}(\tau_1,\tau_2)\bigg]\,
\end{align}
where 
\be
\Xi(\tau_1,\tau_2)=\chi_{p}(\tau_1) \chi_{|-\p+\k|}(\tau_1)\, \chi^{\ast}_{p}(\tau_2) \chi^{\ast}_{|-\p+\k|}(\tau_2)
\ee
and we used the following identity
\be
\int_{-\infty}^0 dt_1 f(t_1)\int_{-\infty}^0 dt_2 f^*(t_2)=2\text{Re}\int_{-\infty}^0 dt_1 f(t_1)\int_{-\infty}^{t_1} dt_2 f^*(t_2).
\ee
In a similar way 
the sum of (b) and (c) diagrams is found as,
\begin{align}\label{P20+P02}
P_2(k)&\equiv P_{02}(k)+P_{20}(k)\\ &=-8iP_{\phi}(k)\int d^3\p \int d{\cal T}_1\ 
\mathrm{Re} \vphi_{k}(\tau_1) \int^{\tau_1} d{\cal T}_2\ 
\bigg[\Xi(\tau_1,\tau_2)- \Xi^{\ast}(\tau_1,\tau_2)\bigg]\text{Im}\vphi_{k} (\tau_2)\nonumber
\end{align}

Putting these all together, the two-point function is therefore
\begin{align}
\nonumber
\expect{\vphi_{\k} \vphi_{-\k}}&=P_{11}(k)+P_{2}(k)
\\
\nonumber
&=8P_{\phi}(k)\int d^3\p \int d{\cal T}_1\ 
\mathrm{Re} \vphi_{k}(\tau_1) \int^{\tau_1} d{\cal T}_2 \ 
\bigg [ \Xi(\tau_1,\tau_2)\vphi^{\ast}_{k} (\tau_2)+\Xi^{\ast}(\tau_1,\tau_2)\vphi_{k} (\tau_2) \bigg]
\\
&=16 P_{\phi}(k)\ {\rm Re} \int d^3\p \int^0_{-\infty} d{\cal T}_1 
\, \text{Re}\vphi_{k}~ \chi_{p}\,\chi_{|-\p+\k|}~
\int^{\tau_1}_{-\infty} d{\cal T}_2\ 
\vphi^*_{k}~ \chi^*_{p}\,\chi^*_{|-\p+\k|}\label{2pf}
\end{align}
It is instructive to compare \eqref{P11} and \eqref{P20+P02}. Besides the sign difference in the $\chi$ dependent factors $\Xi, \Xi^*$, the $\vphi$ dependent parts have a crucial difference: 
In \eqref{P11} we have two $\text{Re}\vphi$ factors while in \eqref{P20+P02} we have one $\text{Re}\vphi$ and one $\text{Im}\vphi$. Recalling \eqref{G-P-prop} and that $\text{Re}\vphi$ is proportional to a propagator, this difference may already be seen directly from the diagrams in Fig. \ref{Power-spectrum:Feyn-Dia--One-loop}, where in (a) we have two solid black lines and in (b) and (c) one solid and one dashed black line. In the next subsection, computing contributions of \eqref{P11} and \eqref{P20+P02} we will observe that \eqref{P11} dominates over \eqref{P20+P02}.

\subsection{One-loop integrals}
Given the expressions we compute the integrals explicitly. While we are interested in the expression for the two point function \eqref{2pf}, it is instructive to present computation of  \eqref{P11} and \eqref{P20+P02} separately. To preform the computations we need to know the mode functions for $\chi$ and $\phi$ fields. For the $\vphi$ field, we use the standard Bunch-Davies vacuum mode function, \eqref{varphi-mode}:
\be\label{phi-Re-Im}
\text{Re} \varphi_{\k}= \dfrac{H}{\sqrt{2 k^3}} (\sin k\tau-k\tau \cos k\tau),\qquad \text{Im} \varphi_{\k}= \dfrac{H}{\sqrt{2 k^3}} (\cos k\tau+k\tau \sin k\tau).
\ee
For the $\chi$ modes, as discussed we need to consider the (narrow) resonance particle production,
\be\label{chi-mode-general}
\chi_{\p}(\tau) \simeq \dfrac{-H \tau}{\sqrt{2p}} \left( e^{-i p \tau}+|\beta_\p| \theta(\dfrac{\alpha}{2}+p\tau) e^{+i p \tau} \right)\simeq \dfrac{-H \tau}{\sqrt{2p}}  e^{-i p \tau},
\ee
where $\theta(\dfrac{\alpha}{2}+p\tau)$ is the condition for the mode with momentum $p$ to be excited by the time $\tau$, and $|\beta_{\p}| = \sinh(q^2 \alpha / 4)\ll 1$. Therefore, 
\be\label{Xi}
\Xi(\tau_1,\tau_2)=\frac{H^4\tau_1^2\tau_2^2}{4p|\k-\p|}\ e^{-i(p+|\k-\p|)(\tau_1-\tau_2)}.
\ee

The details of the integral over time variables $\tau_1,\tau_2$ is given in the appendix \ref{appen:two-pint-one-loop} and here we quote the final result:
\begin{equation}\begin{split}
\expect{\vphi_{\k} \vphi_{-\k}}&=\frac{{\cal G}^2 f^2 P_{\phi}(k)}{2}\  \int \dfrac{d^3\p}{pk|-\p+\k|} \ ({\mathbf{I}_{11}(p,k)}+{\mathbf{I}_{2}(p,k)})\\ 
&= \frac{\pi{\cal G}^2 f^2 P_{\phi}(k)}{4H^2\alpha}\  \int \dfrac{d^3\p}{pk|-\p+\k|}  \left[(\cos\Theta+\frac{\ell}{k\alpha}\sin\Theta)^2+\frac{1}{\alpha^2}\cos^2\Theta+\frac{\ell}{k\alpha}\right]
\\ 
&\simeq \frac{\pi{\cal G}^2 f^2 P_{\phi}(k)}{4H^2\alpha}\  \int \dfrac{d^3\p}{pk|-\p+\k|}  \big(\cos\frac{\alpha k}{\ell}-\frac{\ell}{\alpha k}\sin\frac{\alpha k}{\ell}\big)^2.\label{phi-phi-3}
\end{split}
\end{equation}
In the above, ${\mathbf{I}_{11}(p,k)}, {\mathbf{I}_{2}(p,k)}$ whose explicit expressions are given in \eqref{I-11}, \eqref{I-2}, are respectively contributions from $P_{11}$ and $P_2$ and, 
\be
\ell=p+|\k-\p|,\qquad \Theta=-\frac{\alpha}{2}\ln(\frac{\ell+k}{\ell-k}).
\ee
In the last line in \eqref{phi-phi-3} we used the fact that for $k\lesssim \ell$, $\Theta\simeq -\frac{\alpha k}{\ell}$. Moreover, since $\alpha\gg 1$ one can safely drop the last two terms.  The $k\lesssim \ell$ assumption is justified noting that, as we will see below explicitly, the main contribution to the momentum integrals comes from the subhorizon momenta $p$ while we are only interested in the value of the two-point function for the superhorizon modes $k$ which are relevant to the power spectrum of curvature perturbations. 

Having performed the time integrals, we now focus on momentum integrals,
\be\label{2pf-simp}
\begin{split}
\dfrac{\expect{\vphi_{\k} \vphi_{-\k}}}{ P_{\phi}(k)}&= \frac{\pi{\cal G}^2 f^2 }{4H^2\alpha}\   \int_{p_{_{IR}}}\  \dfrac{p^2dp\ d\Omega_p}{pk|-\p+\k|} \big(\cos\frac{\alpha k}{\ell}-\frac{\ell}{\alpha k}\sin\frac{\alpha k}{\ell}\big)^2
\end{split}
\ee
where as we have argued $p_{_{IR}}=\text{max}(H, A {k})=Ak$, with $A\simeq 1+1/\sqrt{\alpha}\gtrsim 1$. 
For large $\alpha$, $\frac{\alpha}{2u\pm 1}\simeq \frac{\alpha}{2u}\mp \frac{1}{\alpha}(\frac{\alpha}{2u})^2$, and the integral simplifies to,
\be\label{2pf-simp-final}
\begin{split}
\dfrac{\expect{\vphi_{\k} \vphi_{-\k}}}{ P_{\phi}(k)}&=\frac{2\pi^2{\cal G}^2 f^2}{4H^2}\ \int_A^\infty du\ \int_{\frac{\alpha}{2u-1}}^{\frac{\alpha}{2u+1}}\ \frac{dz}{z^2} (\cos z-\frac{1}{z}\sin z)^2\cr
&=\frac{2\pi^2{\cal G}^2 f^2}{4H^2}\ \int_0^{\frac{\alpha}{2A}}  \frac{dz}{z^2} (\cos z-\frac{1}{z}\sin z)^2\cr 
&\simeq \frac{\pi^3}{24} \left(\frac{\mathcal{G}^2 f^2}{H^2}\right):= \frac{\pi^3}{24} \left(\frac{\mathbb{H}}{H}\right)^2.
\end{split}
\ee
The above is our main result of the section and deserves some comments 
\begin{itemize}
    \item In principle $\mathbb{H}$ can be (much) larger than $H$ within a reasonable range of parameters, while the theory is still perturbative (when the coupling ${\cal G}\ll 1$). 
\item The  $k$ dependence in $\expect{\vphi_{\k} \vphi_{-\k}}$, to leading slow-roll corrections, is essentially the same as $P_{\phi}(k)$  
and hence the power spectrum, even if dominated by the above one-loop result, will still be almost scale invariant. 
\end{itemize}
As we will discuss in section \ref{sec:observables} these two points have significant observational implications.

\subsection{Time integral resonances and  localization in momentum integrals}\label{sec:4-3}

Our final result for the two-point function \eqref{2pf-simp-final} is outcome of a one-loop analysis involving two time integrals and a momentum integral. These integrals, as our explicit calculations show, are finite due to presence of time-dependent coupling constant. Here we discuss salient features of each of these integrals and the resonances which happens for each.

\paragraph{Time integral resonances.} The specific feature of our interactions, as discussed and is manifestly seen in \eqref{interaction-Hamiltonian}, is the oscillatory coupling, with the frequency much larger than the Hubble expansion rate, $\alpha\gg1$. That is, in a single e-fold the coupling oscillates many times ($\alpha$ times). This leads to a resonance effect which is made manifest through the integrals \eqref{saddle-point:1}, \eqref{saddle-point:2}.  As these integrals show, the resonance happens at time $-\tau_s$ \eqref{eta-s}, where $p\, \tau_s \sim \alpha \gg 1$. Therefore the dominant contribution to the time integral comes from the modes well below the horizon scale, which are the modes  entering our momentum integrals. 
Moreover, we note that each time integral leads to a $1/\sqrt{\alpha}$ factor. That is, the two time integrals yield a $1/\alpha$ suppression factor. 

Another very important feature of the result of time integrals is that the contribution to the one-loop two-point function coming from diagram (a)  in Fig.\ref{Power-spectrum:Feyn-Dia--One-loop} (\emph{cf.} \eqref{I-11}) is larger than the contribution of (b)+(c) diagrams (\emph{cf.} \eqref{I-2}) by a factor of $\alpha$. That is, the main contribution comes from diagrams which have $\vphi_1$ in it and not $\vphi_2$. Recalling \eqref{varphi-1} we see that it only involves dashed lines of $\chi$ field (contractions of $\chi$-fields and not propagators) whereas $\vphi_2$ \eqref{varphi-2} involves both dashed lines and propagators of $\chi$. This, noting the detailed analysis of the appendix \ref{appen:two-pint-one-loop}, is rooted in the fact that propagators involve commutators of the fields (see \eqref{propagator-phi}), while the dashed lines the symmetric product of the fields. The leading order $\alpha$ contribution cancels out in the anti-symmetrization. 

\begin{figure}[ht]
    \centering
    \includegraphics[scale=0.45]{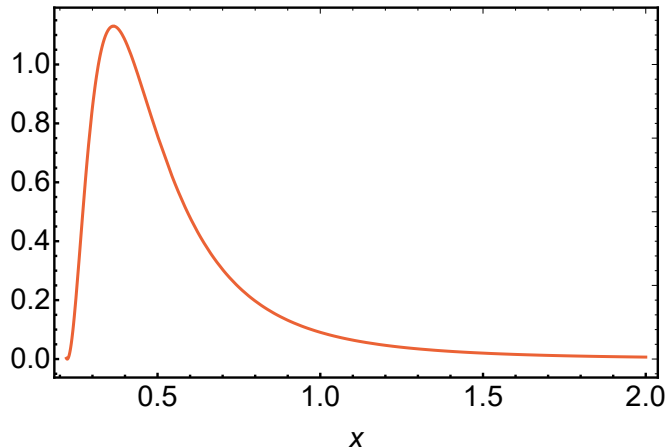}
    \caption{The function $\big(\cos\frac{1}{x}-x\,\sin\frac{1}{x}\big)^2$. In our integrand $x=2p/(\alpha k)$. As we see this function has a sharp peak around $x\sim 0.5$ and falls off  as $1/x^3$ in large $x$.}
    \label{fig:integran-p}
\end{figure}

\paragraph{Localization of integrand in momentum integrals.} Momentum integrals in one-loop two-point function of massless scalar $\varphi^3$ in 4d have logarithmic UV divergence, however, as  \eqref{2pf-simp-final} shows we get a finite result. This happens due to the special form of the integrand in \eqref{2pf-simp}. In the UV $p\gg k$ regime $\ell\simeq 2p$ and the oscillatory part has a sharp peak around $p\sim\alpha k$, as depicted in Fig. \ref{fig:integran-p}. 
That is, the integral receives contribution mainly from the  $p\sim\alpha k$ region. In this region the integrand is proportional to $dp/k$ and we hence end up with an overall enhancement  factor of $\alpha$, which is eventually canceled by the $1/\alpha$ factor coming from time integrals discussed above. That is, in the end the leading contribution to the one-loop two-point function starts from $\alpha^0$, as in the tree level result. There are of course subleading $1/\alpha$ contributions from various different places.
In other words, we not only get a cutoff on momentum, but the integrals are localized around $p\sim\alpha k$  momentum.  
As we will see in the next sections existence of a momentum cutoff is not limited to the one-loop two-point function and is a generic feature of all $L$-loop diagrams in our theory. 

\begin{figure}[ht]
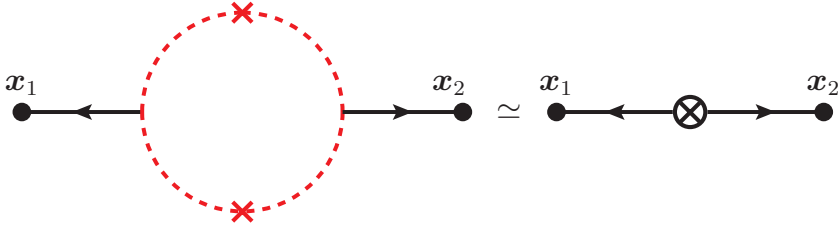

\begin{axopicture}(250,75)(-80,-40)     
    \SetWidth{1.5}
    \SetColor{Black}
    \Vertex(0,0){3.5}
    \ArrowLine(45,0)(0,0)
    \SetColor{Red}
    \DashCArc(82.5,0)(37.5,0,90){3}
    \SetColor{Red}
    \DashCArc(82.5,0)(37.5,90,180){3}
    \SetColor{Red}
    \DashCArc(82.5,0)(37.5,180,360){3}
    \SetColor{Red}
    \SetColor{Black}
    \ArrowLine(120,0)(165,0)
    \Vertex(165,0){3.5}
    \Text(0,10){$\x_1$}
    \Text(160,10){$\x_2$}
    \SetColor{Red}
    \Text(82.5,-37.5){\Cross}
    \Text(82.5,37.5){\Cross}
\end{axopicture}
\begin{axopicture}(0,75)(-80,-40)  
\SetWidth{1.5}
\SetColor{Black}
\Text(-60,0){$\mathbf{\ \ \simeq \qquad} $}
\Vertex(-50,0){3.5}
\Vertex(+50,0){3.5}
\Text(-50,10){$ \x_1$}
\Text(50,10){$\x_2$}
\ArrowLine(-6.5,0)(-50,0)
\ArrowLine(6.5,0)(50,0)
\Text(0,0){\Cross}
\Text(0,0){\OTimes}
\end{axopicture}
\caption{The left figure is graph (a) in Fig.\ref{Power-spectrum:Feyn-Dia--One-loop} which dominates the two point function and the right graph is the depiction of the dominant contribution after performing time and momentum integrals, as we discussed above. The   ``$\chi$-loop effects'' which leads to the one-loop amplification,  captured through the $\left(\frac{\mathbb{H}}{H}\right)^2$ factor in \eqref{2pf-simp-final}, is depicted as the $\boldsymbol{\otimes}$. }\label{Effective-propagator}
\end{figure}

Localization  in momentum integrals can be viewed in the following inspiring way. The integrand in \eqref{2pf-simp} may be written as
\be\label{integrand-P}
\frac{1}{k p|\p -\k|} \big(\cos\frac{\alpha k}{\ell}-\frac{\ell}{\alpha k}\sin\frac{\alpha k}{\ell}\big)^2=\frac{2k^3}{H^2}\frac{2}{H^2\alpha^2}\ (G_\vphi(0,\tau=-\frac{\alpha}{\ell}; \k))^2,
\ee
where we used $p\simeq |\p-\k|\simeq \ell/2$ and $G(0,\tau; \k)$ is the propagator of a mode of momentum $\k$ from $\tau$ to 0, \eqref{G-P-prop}. The important point is that the propagator in \eqref{integrand-P} is computed at the resonance time $\tau=-\alpha/\ell$. This provides an interesting interpretation for the computation here in which there is no explicit appearance of the $\chi$ particles,  an ``effective'' field theory description in which quantum $\chi$-modes are ``integrated out''. This dovetails with the discussions of the previous paragraph, that $\vphi_1$ contributions dominate over the $\vphi_2$ ones, and if $\mathbb{H}\gtrsim H$ it also dominates over the tree level result. In this effective field theory, a $\vphi$ state of momentum $\k$ propagates from $\tau=0$ to $\tau=-\alpha/k$ and then bounces back to $\tau=0$. Or more explicitly, due to $\chi$- fields which are integrated out, the power spectrum at $\k$ is dominated by a $\vphi$ particle pair created at $\tau=-\alpha/k$ which fly off to superhorizon scales in $\tau=0$. These analyses and results may  be summarized in Fig. \ref{Effective-propagator}.
The above interpretation also shows how and why the momentum integral is cutoff (localized): the propagator vanishes for superhorizon modes, i.e. at large $p$ (large $\ell$), $-k \tau= \frac{k\alpha}{\ell}\ll 1$. Therefore, one do not expect any UV divergences.


As another remark, we should note that $G(0,\tau=-\frac{\alpha}{\ell}; \k)$ is propagator of a mode with momentum $\k$, whereas the $\chi$ modes in graph (a) of Fig.\ref{Power-spectrum:Feyn-Dia--One-loop}, have momentum $p$ and $|\p-\k|$. That is, while $\chi$-mode expectation values (and not propagators) are dominating the one-loop result, these $\chi$-modes are  typically created (deep) inside horizon. As such, the $\chi$-modes `running in the loop' are not classical particles produced in (resonant) pair productions.

\paragraph{Terms at the order $\beta_p.$}

As was already mentioned, analysis of section \ref{sec:2-3} indicates that the effects of particle production in the narrow band resonance on the power spectrum is not significant. This is mainly due to the fact that the  Bogoliubov coefficient $\beta_p$ is very small. To see this, let us consider contributions to the $\expect{\varphi\varphi}$ two point function at first order in $\beta_p$. This is achieved by replacing one of the $\chi$ modes contributing to the diagrams in Fig. \ref{Power-spectrum:Feyn-Dia--One-loop} with $\beta_p$ mode in \eqref{chi-mode-expansion-2} rather than the $\alpha_p$ mode. The computations are very similar to those already shown. So, we skip the details and present the final result, which as expected is the order $\beta_p^0$ result with an extra factor of $n_p^{1/2}\simeq q^2\alpha/4$ (\emph{cf.} \eqref{n-K}): 
\be\label{two-point-beta:final}
\begin{split}
\dfrac{\expect{\varphi_{\k} \varphi_{-\k}}_{\mathrm{pp}}}{ P_{\phi}(k)}\sim \frac{\pi^3}{24}\dfrac{\mathcal{G}^2 f^2}{H^2} 
n_p^{1/2} \sim \frac{\pi^3}{24 {\cal G}^2\alpha^3} \left(\dfrac{\mathbb{H}}{H}\right)^6.
\end{split}
\ee
The above expression is smaller than the $\beta^0$ contribution by a factor of $q^2\alpha\sim 10^{-3}$ for viable realizations of our model. There are $\beta_p^l$ contributions $l>1$, which are relatively suppressed by powers $(q^2\alpha)^l$. So, one may safely ignore these contributions. 

\section{Three-point function, one-loop analysis}\label{sec:5}

In this section we  study the three-point function of  scalar perturbations due to  interactions   \eqref{interaction-Hamiltonian}, which is a loop effect, starting from one-loop. We use the general in-in formulation developed in section \ref{sec:3-2} to compute one-loop three-point function of the inflaton fluctuations,
\begin{align}\label{3-pt-one-loop-1}
\nonumber
\hspace{-5mm}\expect{\vphi(\x_1)\vphi(\x_2) \vphi(\x_3)} =&\expect{\vphi_1(\x_1)\vphi_1(\x_2) \vphi_1(\x_3)}
\\
\nonumber
+ &\expect{\vphi_2(\x_1)\vphi_1(\x_2) \vphi_0(\x_3)}+ \expect{\vphi_2(\x_1)\vphi_0(\x_2) \vphi_1(\x_3)}+\expect{\vphi_1(\x_1)\vphi_2(\x_2) \vphi_0(\x_3)}
\\
\nonumber
+ &\expect{\vphi_1(\x_1)\vphi_0(\x_2) \vphi_2(\x_3)}+ \expect{\vphi_0(\x_1)\vphi_2(\x_2) \vphi_1(\x_3)}+\expect{\vphi_0(\x_1)\vphi_1(\x_2) \vphi_2(\x_3)}
\\
+&\expect{\vphi_3(\x_1)\vphi_0(\x_2) \vphi_0(\x_3)} 
+\expect{\vphi_0(\x_1)\vphi_3(\x_2) \vphi_0(\x_3)} 
+\expect{\vphi_0(\x_1)\vphi_0(\x_2) \vphi_3(\x_3)} 
\end{align}
where $\vphi_i$ is the inflaton variation at $i$-order, see section \ref{sec:3-2}.
\begin{figure}[t]
\begin{axopicture}(70,300)(-40,-210)
    \SetWidth{1.5}
    \SetColor{Black}
    \Vertex(82.5,75){3.5}
    \ArrowLine(82.5,37.5)(82.5,75)
    \Vertex(0,0){3.5}
    \ArrowLine(45,0)(0,0)
    \SetColor{Red}
    \DashCArc(82.5,0)(37.5,0,90){3}
    \Text(111.5,24){\Plus}
    \SetColor{Red}
    \DashCArc(82.5,0)(37.5,90,180){3}
    \Text(82.5,-37.5){\Cross}
    \SetColor{Red}
    \DashCArc(82.5,0)(37.5,180,360){3}
    \Text(54.5,24){\Plus}
    \SetColor{Red}
    \SetColor{Black}
    \ArrowLine(120,0)(165,0)
    \Vertex(165,0){3.5}
    \Text(80,-60){(a)}
  \end{axopicture} 
  \begin{axopicture}(70,300)(-200,-210)
    \SetWidth{1.5}
    \SetColor{Black}
    \Vertex(82.5,75){3.5}
    \DashLine(82.5,37.5)(82.5,75){3}
    \Text(82.5,56.25){\Cross}
    \Vertex(0,0){3.5}
    \ArrowLine(45,0)(0,0)
    \SetColor{Red}
    \DashCArc(82.5,0)(37.5,0,90){3}
    \Text(111.5,24){\Plus}
    \SetColor{Red}
    \ArrowArc(82.5,0)(37.5,90,180)
    \SetColor{Red}
    \DashCArc(82.5,0)(37.5,180,360){3}
    \Text(82.5,-37.5){\Cross}
    \SetColor{Red}
    \SetColor{Black}
    \ArrowLine(120,0)(165,0)
    \Vertex(165,0){3.5}
    \Text(80,-60){(b)}
  \end{axopicture} 
  \begin{axopicture}(70,300)(100,-60)
    \SetWidth{1.5}
    \SetColor{Black}
    \Vertex(82.5,75){3.5}
    \DashLine(82.5,37.5)(82.5,75){3}
    \Text(82.5,56.25){\Cross}
    \Vertex(0,0){3.5}
    \ArrowLine(45,0)(0,0)
    \SetColor{Red}
    \ArrowArc(82.5,0)(37.5,0,90)\ArrowArc(82.5,0)(37.5,90,180)
    \SetColor{Red}
    \SetColor{Red}
    \DashCArc(82.5,0)(37.5,180,0){3}
    \Text(82.5,-37.5){\Cross}
    \SetColor{Black}
    \DashLine(120,0)(165,0){3}
    \Text(142.5,0){\Cross}
    \Text(80,-60){(c)}
    \Vertex(165,0){3.5}
  \end{axopicture}
  \begin{axopicture}(70,300)(-60,-60)
    \SetWidth{1.5}
    \SetColor{Black}
    \Vertex(82.5,75){3.5}
    \DashLine(82.5,37.5)(82.5,75){3}
    \Text(82.5,56.25){\Cross}
    \Vertex(0,0){3.5}
    \ArrowLine(45,0)(0,0)
    \SetColor{Red}
    \ArrowArcn(82.5,0)(37.5,360,180)
    \ArrowArc(82.5,0)(37.5,90,180)
    \SetColor{Red}
    \SetColor{Red}
    \DashCArc(82.5,0)(37.5,0,90){3}
    \Text(111.5,24){\Plus}
    \SetColor{Black}
    \DashLine(120,0)(165,0){3}
    \Text(142.5,0){\Cross}
    \Text(80,-60){(d)}
    \Vertex(165,0){3.5}
  \end{axopicture}
  \caption{One-loop contribution to the $\expect{\vphi(\k_1)\vphi(\k_2) \vphi(\k_3)}$ three-point function. Solid lines depict the retarded Green function while dashed lines are expectation value of the free field. Black and red colors are associated with those the $\vphi$ and $\chi$ perturbations respectively. Diagram (a) corresponds to $\expect{\vphi_1(\x_1)\vphi_1(\x_2) \vphi_1(\x_3)}$, diagram (b) to 2nd and 3rd lines in \eqref{3-pt-one-loop-1} and diagrams (c) and (d) to the last line in \eqref{3-pt-one-loop-1}. }
  \label{Bispectrum:Feyn-Dia-one-loop}
  \end{figure}
  

1-PI Feynman diagrams for this one-loop three-point function  are depicted in Fig. \ref{Bispectrum:Feyn-Dia-one-loop}. To show the essence of the computation, we present details of computation of the diagram (a), whose amplitude $B_{111}$ is,
\begin{align}
B_{111}(k_1,k_2,k_3) & \equiv \expect{\vphi_{1}(\k_1)\vphi_{1}(\k_2)\vphi_{1}(\k_3)}\nonumber
\\
\nonumber
&=  8 P_{\phi}^{1/2}(k_1)P_{\phi}^{1/2}(k_2)P_{\phi}^{1/2}(k_3)\int 
d{\cal T}_1\, d{\cal T}_2\,d{\cal T}_3\  \mathrm{Re} \vphi_{\k_1}(\tau_1) \mathrm{Re} \vphi_{\k_2}(\tau_2) \mathrm{Re} \vphi_{\k_3}(\tau_3)\times
\\
&~~~~~~~~~~~~~~~~~~~~~~~~~~~~~~~~~~~~~~~~~\big\langle (\hat{\chi}^2)_{\k_1}(\tau_1) (\hat{\chi}^2)_{\k_2}(\tau_2) (\hat{\chi}^2)_{\k_3}(\tau_3)\big\rangle
\end{align}
where  we have used \eqref{G-P-prop} and \eqref{d-cal-T}, i.e., 
$$
\int d {\cal T}_i\equiv \int \frac{d \tau_i}{ H^4 \tau_i^4} g_3(\tau_i).$$ 
Moreover the term in the brackets above is found to be 
\begin{align}
\nonumber
\big\langle
(\hat{\chi}^2)_{\k_1}(\tau_1) (\hat{\chi}^2)_{\k_2}(\tau_2) (\hat{\chi}^2)_{\k_3}(\tau_3)\big\rangle &= 2\int d^3\p ~\bigg[\,\chi_{\p}(\tau_1) \chi_{-\p+\k_1}(\tau_1)\,\chi_{\p+\k_2}(\tau_2)\chi^{\ast}_{-\p}(\tau_2) \, \chi^{\ast}_{\p-\k_1}(\tau_3)\chi^{\ast}_{-\p-\k_2}(\tau_3)
\\
&+\chi_{\p}(\tau_1) \chi_{-\p+\k_1}(\tau_1)\,\chi_{-\p-\k_3}(\tau_2)\chi^{\ast}_{\p-\k_1}(\tau_2) \, \chi^{\ast}_{-\p}(\tau_3)\chi^{\ast}_{\p+\k_3}(\tau_3) \bigg]
\end{align}

Noting that, for $p \tau \gg 1$ 
\be
\nonumber
\chi_{\p}(\tau)=\dfrac{-H \tau}{\sqrt{2p}}e^{-i p \tau},
\ee
the contribution of diagram (a) to the three-point function can be written as 
\begin{align}
\nonumber
\hspace{-5mm}B_{111}(k_1,k_2,k_3) = & 2\, P_{\phi}^{1/2}(k_1)P_{\phi}^{1/2}(k_2)P_{\phi}^{1/2}(k_3) \int 
\dfrac{d^3 \p}{p |\p-\k_1||\p+\k_2|}
\int d{\cal T}_1  d{\cal T}_2 d{\cal T}_3 \,  (H^2\tau_1^2)(H^2\tau_2^2)(H^2\tau_3^2)\ 
\nonumber
\\
&\ \mathrm{Re} \vphi_{\k_1}(\tau_1)\ \mathrm{Re} \vphi_{\k_2}(\tau_2)\, \mathrm{Re} \vphi_{\k_3}(\tau_3)\, 
e^{ -i (p +|\p-\k_1|)\, \tau_1}\ e^{i(p -|\p+\k_2|)\tau_2}\ e^{i(|\p-\k_1|+|\p+\k_2|)\tau_3}. 
\end{align}
To proceed further we need to perform time intergals. As in the previous section the latter can be computed in the stationary phase approximation where as before at the resonance time $\tau_s$ (saddle point time) $k_i\tau_s\gtrsim 1$. This latter, as discussed, yields a UV cutoff on the loop momentum integral, $p\lesssim \alpha k$, where $k$ is a typical value of the external momenta $k_i$. Using \eqref{saddle-point:1} we obtain
\begin{align}
\nonumber
B_{111}(k_1,k_2,k_3)&=  -i\dfrac{{\cal G}^3 f^3}{512\sqrt{2}} \dfrac{1}{k_1^2 k_2^2 k_3^2} \sqrt{\dfrac{\pi^3}{\alpha^3}}
\int^{\alpha k} \dfrac{d^3 \p}{p \,|-\p+\k_1| |\p+\k_2|}
\\
\nonumber
&\left(e^{-i \Phi \left(|\p-k_1|-k_1+p\right)}+e^{-i \Phi \left(|\p-k_1|+k_1+p\right)}\right)
\left(e^{+i \Phi \left(-|\p+k_2|+k_2+p\right)}-e^{-i \Phi \left(|\p+k_2|+k_2-p\right)}\right)\times
\\
&\left(e^{i \Phi \left( |\p-\k_1|+|\p+\k_2|-k_3 \right)}+e^{i \Phi \left(|\p-\k_1|+|\p+\k_2|+ k_3 \right)}\right)
\label{B111-time-int:1}
\end{align}
where we used  $P_{\phi}(k)=H^2/2k^3$ and $\Phi(k)$ is given in \eqref{Phi},
\begin{align}
\nonumber
    \Phi(k) \equiv \alpha \log \alpha - \alpha -\alpha \log k +\pi/4.
\end{align}
As we see, each time integral gives rise to a parametric suppression factor $1/\sqrt{\alpha}$ and a fast oscillating function $\exp (i\alpha \log \tau_s/\tau_{\ast})$, with $\tau_s$ given in \eqref{eta-s}. 

We should next evaluate momentum integrals. These would have been logarithmically divergent if it were not for the outgoing propagators which yield an effective UV cutoff. The integrals simplify to
\begin{align}
B_{111}(k_1,k_2,k_3)=  \dfrac{{\cal G}^3 f^3}{64\sqrt{2}} \left(\dfrac{\pi}{\alpha} \right)^{3/2} \dfrac{1}{k_1^2 k_2^2 k_3^2}\ 
\int^{\alpha k} \dfrac{d p}{p} \int d \Omega_p \big[\sin\Phi(k_{2-},\alpha)+\sin\Phi(k_{2+},\alpha)\big]  
\label{B111a:I}
\end{align}
with
\begin{align}
k_{i \pm}:= k_i \pm \k_i\cdot\hat{\p}\,.f
\end{align}

The other diagrams in Fig. \ref{Bispectrum:Feyn-Dia-one-loop} may be computed in a similar way, some of the details of which may be found in the appendix \ref{Appen:bispectrum-full}. Here we present the final result for the
three-point function at one-loop order $B_{1- \mathrm{loop}}$,
\begin{align}\label{B-one-loop}
 B_{1-\mathrm{loop}}= -i\frac{  f^3 {\cal G}^3}{16 \sqrt{2}}\dfrac{1}{ k_1^2 k_2^2 k_3^2} \left(\frac{\pi}{\alpha }\right)^{3/2}
\int^{\alpha k}\dfrac{d p}{p} \int d\Omega_{\p} \left[ +e^{+i \Phi (k_{2+},\alpha)}-e^{-i \Phi (k_{2-},\alpha)}
\right].   
\end{align}
After a tedious but straightforward calculation\footnote{A Mathemtica code for the calculations is available online from  \href{http://abolhasani.physics.sharif.edu/files/calcs/Bispectrum_Calculator_Function_v4.nb}{here}.}
we get    
\be\label{bispectrum-oneloop}
B_{\varphi}(\k_1,\k_2,\k_3)=\dfrac{2 \pi^{5/2}}{2 \sqrt{2}} \dfrac{{\cal G}^3 f^3}{H^3}\dfrac{1}{k_1^2\, k_2^2\, k_3^2} \dfrac{\ln\alpha}{\alpha^{5/2}} \sin \Phi(2k_2,\alpha).
\ee

As we see in \eqref{B-one-loop} the logarithmic divergence of the momentum integral is tamed by the oscillatory phase factors. As in the two-point function case (\emph{cf}. discussions in section \ref{sec:4-3}), this momentum integral is again localized around $p\sim \alpha k$ momenta.
We will discuss physical significance of the above in section \ref{sec:NGy}. 

\begin{figure}[t]
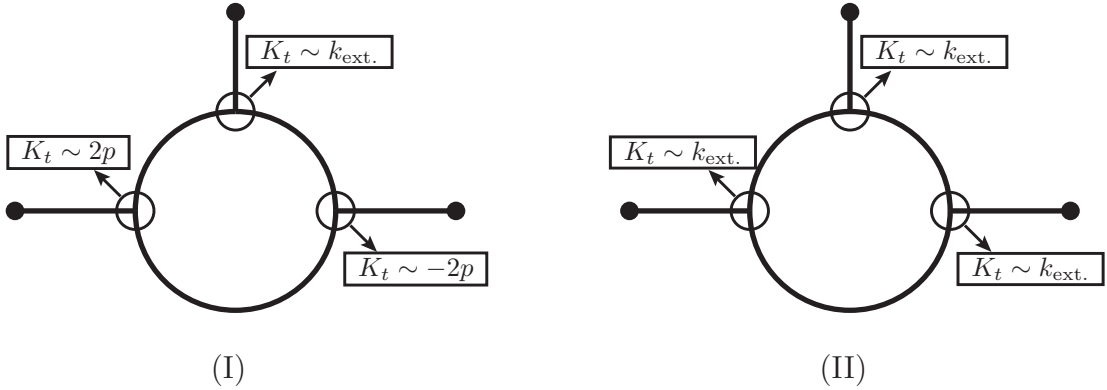

\begin{axopicture}(70,150)(-40,-60)
    \SetWidth{2}
    \SetColor{Black}
    \Vertex(82.5,75){3.5}
    \Line(82.5,37.5)(82.5,75)
    \Vertex(0,0){3.5}
    \Line(45,0)(0,0)
    \Arc(82.5,0)(37.5,0,90)
    \Arc(82.5,0)(37.5,90,180)
    \Arc(82.5,0)(37.5,180,360)
    \SetColor{Black}
    \Line(120,0)(165,0)
    \Vertex(165,0){3.5}
    \SetWidth{1.1}
    \ECirc(45,0){7}
    \LongArrow(39.5,5.5)(31.75,13.25)
    \ECirc(82.5,37.5){7}
    \LongArrow(88,43)(95.75,50.75)
    \ECirc(120,0){7}
    \LongArrow(125.5,-5.5)(133.25,-13.25)
    \BText(20,22){$K_t \sim 2p$}
    \BText(150,-22){$K_t \sim -2p$}
    \BText(114,60){$K_t \sim k_{\mathrm{ext.}}$}
    \Text(80,-60){(I)}
  \end{axopicture} 
  \begin{axopicture}(70,150)(-200,-60)
    \SetWidth{2}
    \SetColor{Black}
    \Vertex(82.5,75){3.5}
    \Line(82.5,37.5)(82.5,75)
    \Vertex(0,0){3.5}
    \Line(45,0)(0,0)
    \Arc(82.5,0)(37.5,0,90)
    \Arc(82.5,0)(37.5,90,180)
    \Arc(82.5,0)(37.5,180,360)
    \SetColor{Black}
    \Line(120,0)(165,0)
    \Vertex(165,0){3.5}
    \SetWidth{1.1}
    \ECirc(45,0){7}
    \LongArrow(39.5,5.5)(31.75,13.25)
    \ECirc(82.5,37.5){7}
    \LongArrow(88,43)(95.75,50.75)
    \ECirc(120,0){7}
    \LongArrow(125.5,-5.5)(133.25,-13.25)
    \BText(20,22){$K_t \sim k_{\mathrm{ext.}}$}
    \BText(150,-22){$K_t \sim k_{\mathrm{ext.}}$}
    \BText(114,60){$K_t \sim k_{\mathrm{ext.}}$}
    \Text(80,-60){(II)}
  \end{axopicture}
 \caption{The two leading order possibilities for the momentum $K_t$ at each vertex, \emph{cf.} \eqref{kt-3-pt-funcn}.}
  \label{3point-1loop-kt:fig}
  \end{figure}

In the computations of previous section we found resonance in time integrals and  localization in momentum integrals. We close this section by remarks on similar features in the one-loop three-point function. In this case we deal with time integrals which are typically of the form
\begin{align}\label{kt-3-pt-funcn}
\int^{\tau} \dfrac{d \tau'}{\tau'} \sin\omega t'\, e^{\pm i |K_t| \tau'} \propto e^{\pm i\Phi(|K_t|,\alpha)},
\end{align}
where $K_t$ 
is a function of external momenta $\k_1$,$\k_2$,$\k_3$ and the loop momentum  $\p$, and can be positive or negative. In the  $p\gg k_i$ limit  $K_t$ can be  either $K_t\sim\pm 2p$ or   something of the order of external momenta $K_t\sim k_i$, respectively depending on whether a product of $\chi \chi$ and $\chi^{\ast} \chi^{\ast}$ or  $\chi \chi^{\ast}$ appears at a vertex. This leaves us with some different possibilities for $K_t$ where  at each of the three vertices  $K_t\sim 2p $, $K_t\sim -2p $ or $K_t\sim k_{\mathrm{ext.}}$. Detailed analysis of the one-loop three-point function (see, \eqref{B111a:I},\eqref{B210b:I},\eqref{B300c:I} and \eqref{B300d:I}) reveals that only the \emph{two} possibilities depicted in Fig. \ref{3point-1loop-kt:fig} happen: Either we get a pair of vertices with $K_t\sim 2p$ and $K_t\sim -2p$ while the last one has $K_t \sim k_{\mathrm{ext}}$ or  all vertices  have $K_t \sim k_{\mathrm{ext}}$.

With this argument let us now explore the phase $\Phi(K_t)$ in \eqref{kt-3-pt-funcn}. 
For the case (I) phases of two vertices  with $K_t \sim \pm 2p$ cancel out and we end up with a single $\exp(\pm i\Phi(\k_i))$. For  case (II) it happens that all $p$ dependence cancels at vertices and we remain with a $\exp(\pm i\Phi(\k_1)\pm i\Phi(\k_2)\pm i\Phi(\k_3))$ phase from time integrations.  Remarkably as shown in the appendix \ref{Appen:bispectrum-full} these diagrams exactly cancel among diagrams $(c)$ and $(d)$. This hence yields momentum localization in one-loop three-point function analysis. Similar feature, as we will argue next, appears in higher-loop order analysis, establishing the existence of  momentum cutoffs  at all orders.

\section{Higher loop calculations, general formulation and analysis}\label{sec:6}
We pointed out in  section \ref{sec:4-3} that the one-loop two-point function result can become comparable or larger than the tree level expression  even in the regime where the coupling is parametrically small. This prompts the question whether similar amplifications can occur in higher loops and hence effectively yielding a break down of perturbation theory. In this section we establish that  suppression by powers of the coupling dominates over amplification effects in higher loops, confirming validity of the perturbative expansion.

\subsection{Generic higher loops and superficial degree  of divergence}\label{sec:6-1} 

We start our higher loop analysis by a review of basics of loop calculations in quantum field theories in general and in the in-in formulation in particular. 
The machinery used in this section is based on the formulation developed in \cite{Musso:2006pt}, generalized and extended to suit our model. 
Let us provide some useful remarks before going on with the calculations of higher loop diagrams. Consider an $n$-point function in  $V$-th order in perturbation theory, $\expect{{\hat O}_n}$ \eqref{O-N:exp}.  For our cubic interaction  ${\cal L}_{int}\propto \vphi \chi^2$, the loop number $L$ (the number of independent \emph{internal} momenta over which we integrate) is then 
\be
\label{no-of-loops}
L= \dfrac{V-E}{2}+1
\ee
where $V$ and $E$ denote the number of vertices and externals legs respectively. Accordingly, for the number of internal lines (loop propagators) $P$, we obtain
\be
\label{no-of-lines}
P=\dfrac{3V-E}{2}.
\ee
See appendix \ref{appen:loop-insertion} for more discussions on generic 1-PI loop diagrams. 

\textbf{Time integrals.} Each interaction vertex comes with a time integral can be calculated using the stationary phase approximation. As shown in the appendix \ref{appen:two-pint-one-loop} each time integral yields a $\alpha^{-1/2}$ suppression and a momentum dependent oscillatory phase factor.  The result depends on the \emph{net energy transfer}  $\Delta E$ 
through a highly oscillating sine function  $\sin (\alpha-\alpha \log \alpha -\alpha \log \tau_i/\tau_{\ast} +\pi/4)$ where $\tau_i$ is the stationary point $\tau_i=-\alpha/|\Delta E|$ and $\tau_{\ast}$ denotes some fiducial time, \emph{cf.} comments below \eqref{Phi}.  This oscillatory phase renders the momentum integrals finite, as we will discuss next.

To summarize, for    the ${\cal G}, f$ and $\alpha$ dependence of a generic $L$-loop $E$-point function  amplitude we find
\be\label{general-estimate}
{P_{E,L}}\sim \left(\frac{{\cal G}^2 f^2}{H^2\alpha} \right)^{L+\frac{E}{2}-1} \ F_{E,L}(\alpha),
\ee
where the first factor comes because of contribution of $V$ couplings, powers of $\alpha, H$ are due to time integrals and $F_{E,L}(\alpha)$ is due to momentum integrals, e.g. for one-loop two-point function $F_{2,1}(\alpha)\sim \alpha$ and for one-loop three-point function $F_{3,1}(\alpha)\sim \log\alpha/\alpha$.
The main remaining computation is hence finding $F_{E,L}(\alpha)$, which we take on next.

\subsection{Momentum integrals and the cutoff}\label{sec:localization}  

To compute  $F_{E,L}(\alpha)$ in \eqref{general-estimate} we should explore more closely the momentum integrals. As we have already seen in some different examples in sections \ref{sec:4}, \ref{sec:5} and appendices \ref{appen:two-pint-one-loop}, \ref{Appen:bispectrum-full}, momentum integral in our case do not lead to divergences. 
Here we would like to establish that this is a generic feature in our model. Since we have massless states running in the loops we need to worry about both IR and UV divergences. 

\paragraph{Absence of IR divergences.} Dealing with massless states running in the loops one should in principle worry about IR divergence. However, due to the fact that we are on an almost de Sitter space and that the propagators vanish for superhorizon scales, there is a natural IR cutoff, the Hubble scale $H$. To be more precise,  consider the retarded Green's function of a massless field with momentum $k$ \eqref{propagator-phi}.\footnote{Note that effective mass of the $\chi$ field is given in \eqref{chi:Mass}. However, as discussed in section \ref{sec:2-3}, the effects of the oscillatory part in the mass term is only pronounced in the short time span of the narrow band resonance, which is captured in $\beta_p$ term effects in our analysis. So, for our discussions here the mass of $\chi$ is essentially $\mu^2$ which we take it to be much smaller than $H^2$.} $G(k;\tau,\tau+\Delta\tau)$  vanishes as $k\Delta\tau (k\tau)^2 P_\phi (k)$  when $k \Delta \tau \ll 1$, as implied by causality. Particularly, when the end points of vertices connected by this propagator are at $\tau_{s_{\pm}}=\alpha/(p\pm k)$, it would vanish as $(k/p)^2$ for $p \gg k$.
 
\paragraph{Superficial degree of UV divergence in momentum integrals.} To warm up, consider a generic $L$-loop 1-PI diagram of this model. It involves $\prod_{l=1}^L (d^3\p_l)$ and product of propagators which in the UV behave as $\p^P$ where $\p$ is a typical loop momentum. Therefore, the superficial degree of UV divergence is 
\be
D=3L-P=3-E. 
\ee
So, for any loop $L$, naively the two and three point functions can be respectively linearly or logarithmically divergent and the higher point functions are just finite. This naive counting should be corrected due to the time integrals, which as we will discuss renders the integrals finite. Moreover, in our analysis we need to go beyond the 1-PI diagrams to establish validity of perturbative expansion beyond one-loop. 

\begin{figure}[H]
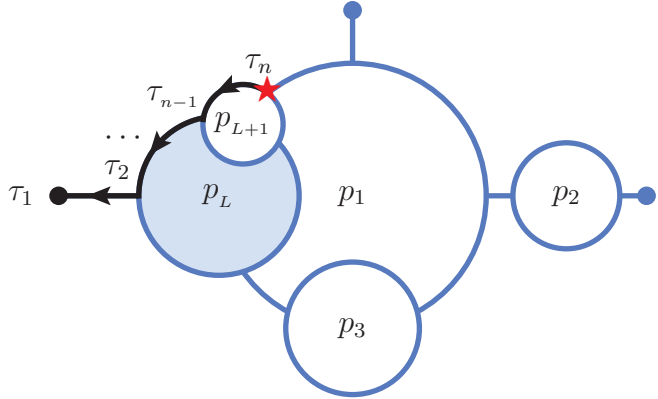

    \begin{axopicture}(70,160)(-160,-77.5)
    \SetWidth{2}
    \SetColor{LightBlue}
    \Vertex(220,0){3.5}
    \Vertex(110,70){3.5}
    \Line(60,0)(0,0)
    \Line(160,0)(220,0)
    \Line(110,50)(110,70)
    \CCirc(110,0){50}{LightBlue}{White}
    \CCirc(60,0){30}{LightBlue}{VeryLightBlue}
    \CCirc(110,-50){25}{LightBlue}{White}
    \CCirc(190,0){20}{LightBlue}{White}
    \CCirc(69,27){15}{LightBlue}{White}
    \SetColor{Black}
    \ArrowLine(30,0)(0,0)
    \ArrowArc(60,0)(30,100,182)
    \ArrowArc(69,27)(15,55,177)
    \Vertex(0,0){3.5}
    \Text(110,0){$p_1$}
    \Text(60,0){$p_{_L}$}
    \Text(110,-50){$p_3$}
    \Text(190,0){$p_2$}
    \Text(69,27){$p_{_{L+1}}$}
    \Text(-14,0){$\tau_1$}
    \Text(22,10){$\tau_2$}
    \Text(25,22){$\cdots$}
    \Text(43,37){$\tau_{_{n-1}}$}
    \Text(75,50){$\tau_n$}
    \SetColor{Red}
    \Text(78,40){$\bigstar$}
  \end{axopicture}
\caption{A general $L+1$ loop diagram. Schematically, the solid arrow line show how starting from an arbitrary vertex one can get to an external leg via a series of head-to-tail \emph{propagators}.}
\label{L+1-loop-route:fig}
\end{figure}

\paragraph{Momentum integrals are naturally cutoff.} As we discussed time integrals yield momentum dependent oscillatory phases and step-functions in time, \emph{cf.} \eqref{saddle-point:1}, \eqref{saddle-point:2}, which  tame the UV behavior of momentum integrals. In the one-loop two and three point functions this led to  UV momentum cutoff $p\sim k\alpha$, where $k$ is a typical (superhorizon) momentum for external legs (note that we are working in the $\alpha\gg 1$ limit).
Recalling our discussion on superficial degree of divergence, this already shows why in the two-point function $F_{2,1}\sim \alpha$ (a ``linear divergence'') and in three-point function $F_{3,1}\sim \log\alpha/\alpha$ (a ``log divergence''). Note that in the case of three-point function the extra suppression by $1/\alpha$ comes from cancellation between the two oscillatory phases in \eqref{B-one-loop}. In what follows we argue, using mathematical induction,  that similar localization happens for a generic loop integral, both 1-PI and non 1-PI diagrams, as depicted in Fig. \ref{L+1-loop-route:fig}.

Before coming to the reasoning,  assuming the validity of the above momentum localization/cutoff argument, we can already write an improved version of \eqref{general-estimate},
\be\label{general-estimate-improved}
{P_{E,L}}\sim \left(\frac{{\cal G}^2 f^2}{H^2\alpha} \right)^{L+\frac{E}{2}-1} \ \alpha^{N_{E,L}},\qquad N_{E,L}\leq D=3-E.
\ee
The above establishes our claim about the validity of perturbation theory when the ``effective coupling'' $\frac{{\cal G}^2 f^2}{H^2\alpha}\ll 1$, provided that we can convincingly argue that $N_{E,L+1}\leq N_{E,L}\leq 3-E$. For the special case of two-point function, i.e. for $E=2$, we have shown $N_{2,1}=1$, already saturating its upper bound. Therefore, the higher loop results for $E=2$ can never dominate over the one-loop and we already have the desired.


Let us now present our inductive argument why none of the loop momenta can exceed $\alpha k$, namely there is a cut on every internal momenta. Consider an arbitrary $L$-loop diagram whose loop momenta $p_1,...,p_L$ obey the following condition 
\be
p_1,\cdots,p_L \lesssim \alpha k,
\ee
where $k$ is a typical momentum for external legs. Inserting a new loop with momentum $P_{L+1}$ as described in the appendix \ref{appen:loop-insertion}, we show that  $p_{L+1} \lesssim \alpha k$. As we argue this is a direct consequence of Heaviside step function in retarded Green's function. First, note that all vertices are reachable from an external leg with at least one \emph{solid} line which is a series of head-to-tail propagators. Let us tag them by $\tau_i$ with increasing indices which start from a external leg at $\tau \rightarrow0 $, see Fig. \ref{L+1-loop-route:fig}. 
We hence have
 \be\label{tau-ordering}
 \tau_1 \geq \tau_2 \geq \cdots \geq \tau_{n-1} \geq \tau_n.
 \ee

As discussed in details, there is  a net energy transfer at each vertex $K_t$ from and to the background (see discussions below \eqref{Phi}) so that at $K_t \tau_s =\pm \omega/H$ we have a saddle point which gives the dominant contribution to the time integrals. Then, \eqref{tau-ordering} implies
  \be
  \label{Kn-order:eq}
 |K_{t_2}| \geq |K_{t_3}| \geq \cdots \geq |K_{t_{L-1}}| \geq |K_{t_L}|.
 \ee
On the other hand, we learnt from detailed calculations of two and three point functions that in the limit that loop momentum pretty much exceeds the external momenta (i) A loop with two external legs both  vertices have non-negligible energy transfer, in particular $ K_t \sim 2p$ (\emph{cf.} \eqref{Xi}) and (ii) A loop with three external legs has exactly \emph{two  vertices} with non-negligible energy transfer (see Fig. \ref{3point-1loop-kt:fig} and the discussions in the end of section \ref{sec:5} and analysis of the appendix \ref{Appen:bispectrum-full}). Therefore, \eqref{Kn-order:eq} yields
\be
 |K_{t_L}| \leq\cdots \leq |K_{t_2}| \lesssim \alpha\, k_{\mathrm{ext.}}.
 \ee

We continue with the proof by contradiction.  Let us pick the starred vertex of the annexed $(L+1)^{th}$ in Fig. \ref{L+1-loop-route:fig}, and let $K_{t_{L+1}} \sim 2 p_{L+1}$.  Therefore, if $p_{L+1} \gtrsim p_{L}$, we get $|K_{t_L}|>|K_{t_{L+1}}|$ which is in contradiction with \eqref{Kn-order:eq}. As a result, we find that $p_{L+1}$ cannot exceeds $p_L$. Therefore, if the loop momenta in $L$-loops of a generic $L+1$ loop diagram are less than $\alpha k$, the last loop momentum  is also less than $\alpha k$, $p_{L+1} \lesssim \alpha k$. Finally, we note that for the outermost loop, those involving a vertex connected to an external leg,  the reasoning is similar to the one-loop case. If this vertex has  $k_t\sim p_1$,  then $p_1\sim\alpha k$. Note that this argument applies to both 1-PI and non-1-PI diagrams. 
This proves our claim that in a general diagram are loop momenta are cut at $p_i \lesssim \alpha k$.

{We close this section by the remark that the step-function part in the time integrals yields momentum cutoffs and the momentum dependent phase factors lead to further suppression factors. In the one-loop two-point function this suppression factor is of order one, while in the one-loop three-point function there there is an extra $1/\alpha$. So, our momentum cut arguments yields an upper bound for the loop integrals, i.e. an upper bound on the power $N_{E,L}$ in \eqref{general-estimate-improved}.}

\section{Observables of the model}\label{sec:observables}

Having established the all loops field theory results for superhorizon two and three point functions, we can  compute the power spectrum of curvature perturbations, its tilt and non-Gaussianty in our model. Then comparing to Planck results we can find physically relevant range of parameters.

\subsection{Scalar to tensor ratio and spectral tilt of curvature  perturbations}\label{sec:7-1}

\paragraph{Power spectrum of curvature perturbations.} As we argued the higher loop two-point function is parametrically smaller than the one-loop one. Therefore, the two-point function in leading order in $\alpha$ is the sum of tree level and one-loop results. Recalling \eqref{2pf-simp-final} we hence have
\be\label{PR-our-result}
{\cal P}_{\cal R}(k) = \dfrac{H^2}{8 \pi^2 M^2_{\text{Pl}}\epsilon}\left[1+ \dfrac{\pi^3}{24}\left( \dfrac{\mathbb H}{H}\right)^2\right]=\frac{\pi{\cal G}^2}{96\alpha^2}\left[1+ \dfrac{24}{\pi^3} \left(\dfrac{H}{\mathbb H}\right)^2\right]
\ee
in which we used the definition of comoving curvature perturbations ${\cal R}(\k) = (H/\dot{\phi}) \varphi(k)$ and \eqref{alpha-H/w}. Moreover, 
we have ignored the back-reaction of created $\chi$ particles, which as argued is small at tree level and remains small at loop levels. Matching with the observations we need to equate the above with the Planck result \cite{Planck-2018}, $P_{\cal R}=2.22\times 10^{-9}$. 

As a direct consequence of this result  the amplitude of scalar  and tensor perturbations are determined by two distinct scales. Owing to this fact the tensor to scalar ratio would be suppressed compared to the usual Lyth bound \cite{Lyth-bound}:
\be
r= \dfrac{{\cal P}_T}{{\cal P}_{\cal R}} = \dfrac{16\, \epsilon}{1+\dfrac{\pi^3}{24}\left( \dfrac{\mathbb H}{H}\right)^2} <16 \,\epsilon
\ee
This suppression is the key feature of this model which allows for convex models of inflation to meet the stringent bounds on the tensor to scalar ratio $r$. 

\paragraph{Spectral tilt.} As we will find in the range of parameters  preferred by the data $\mathbb{H}\gtrsim H$ and the one-loop result in $P_{\cal R}$ dominates. Therefore,  the spectral tilt is 
\be\label{our-tilt}
n_s-1=\dfrac{d\, \ln {\cal P}_s(k)}{d \,\ln k} = 2\, \dfrac{d\, \ln {\cal G}}{d \,\ln k} - \dfrac{d\, \ln {\epsilon}}{d \,\ln k}
= 2 \eta -4\epsilon + 2\, \dfrac{d \ln {\cal G}(\gamma)}{d N},
\ee 
where  $\gamma$ is defined in \eqref{gamma-def} and we used the fact that $\frac{d\, \ln {\epsilon}}{d \,\ln k} = 4\epsilon-2\eta$. Recall that for single field slow-roll models of inflation, the spectral tilt is $n_s-1=2\eta -6 \epsilon$, and it would be handy to  parametrize the last term in \eqref{our-tilt} in terms of a parameter $\beta$, 
\be\label{G-n}
 \beta:=-2\, \dfrac{d \ln {\cal G}(\gamma)}{\epsilon\ d N}.
\ee
Then, reading $\beta$ from the data we can solve \eqref{G-n} for ${\cal G} (\gamma)$.

To see how it works, let us consider a single field model with potential $V(\phi)$ and recall that 
\be
\epsilon= \dfrac{M_p^2}{2} \left(\dfrac{V_{,\phi}}{V}\right)^2,\qquad dN = -M_p^{-2} \dfrac{V}{V_{,\phi}}d \phi.
\ee
Then, \eqref{G-n} becomes
\be
d \ln {\cal G}(\gamma) = \dfrac{\beta}{2} \dfrac{V_{,\phi}\,d\phi}{V}
\ee
and hence,
\be
{\cal G}(\gamma) = {\cal G}(\gamma_i) \left(\frac{V(\phi(\gamma))}{V(\phi(\gamma_i))}\right)^{\beta/2}.
\ee
For the special case of chaotic inflation models, $V(\phi)=\lambda_p \phi^p$, we have
\be
\gamma=\dfrac{\dot{\phi}}{f^2}= \dfrac{M_p}{\sqrt{3} \,f^2}\lambda_p^{1/2} \phi^{\frac{p}{2}-1},
\ee
therefore
\be
V(\phi)= \lambda_p^{\frac{-2}{p-2}}\left(\dfrac{\sqrt{3}f^2}{M_p} \right)^{\frac{2p}{p-2}}~ \gamma^{\frac{2p}{p-2}},\qquad p\neq 2.
\ee
Finally we have ${\cal G}(\gamma) \propto \gamma^{\beta\,p/(p-2)}$.
\begin{figure}[t]
    \centering
    \includegraphics[scale= 0.65]{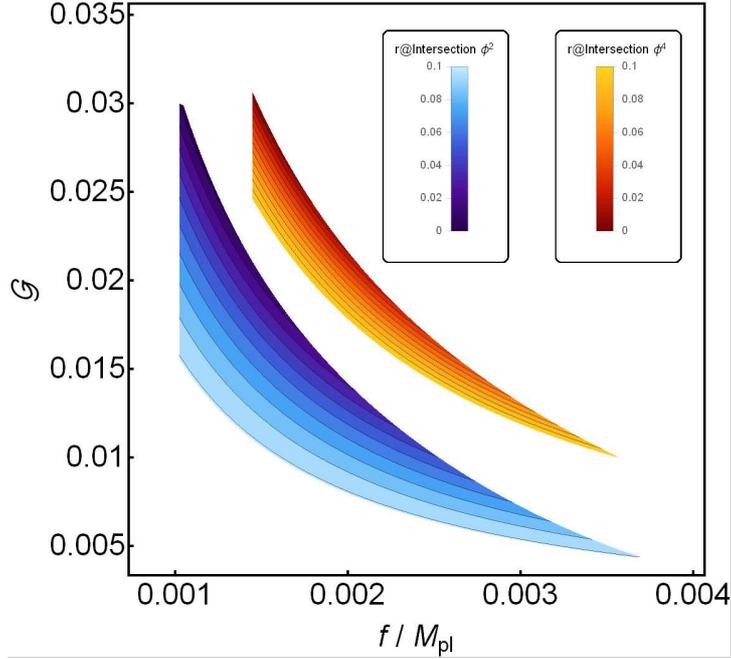}
\caption{Range of parameter allowed by the current observations for $\phi^2$ and $\phi^4$ inflationary models. For this fit, we have used  $P_{\cal R}=2.22\times 10^{-9}$ and  $r<0.1$. We have  assumed $30<\alpha<125$ \cite{Resonant-Monodromy, Planck-2018, omega-bound} and $q<0.1$ to have a reliable perturbative description. The vertical cut on the left comes from $\alpha=125$.}
    \label{fig:parameter-space}
\end{figure}
\paragraph{Comparison with the data.} We can identify the range of allowed parameters by comparing our results with the $P_{\cal R}; n_s, r$ data from Planck and the requirement that we are in a perturbative regime of our model. We first use \eqref{slow-roll-parameters} and \eqref{Ne-epsilon} to relate number of e-folds $N_e$ to $\alpha, f$:
\be
\left(\frac{\alpha}{100}\right)^2\left(\frac{f}{M_{Pl}}\cdot 10^3\right)^2=p\left(\frac{50}{N_e}\right),
\ee
where $p$ is the power of the polynomial potential of the inflaton. In our examples we take $p=2,4$. So, an upper bound on $\alpha$, $\alpha=125$, yields a lower bound on $f$, with $N_e=60$. Then we use \eqref{PR-our-result} to relate ${\cal G}/\alpha$ and $\mathbb{H}/H$. We take $q<0.1$ and recall that
\be
q=\frac{2 p}{\pi^2 N_e P_{\cal R}} \frac{{\cal G}}{r\alpha^4}\simeq 0.002 p \left(\frac{60}{N_e}\right)\left(\frac{{\cal G}}{0.01}\right)\left(\frac{0.07}{r}\right)\left(\frac{100}{\alpha}\right)^4.
\ee
The Planck bound on spectral tilt $n_s$, recalling that 
\be
n_s-1=-\frac{1}{2N_e}\left[p\,(\beta+1)+1\right],
\ee
can then be used to bound parameter $\beta$ \eqref{G-n}. For $1-n_s\sim few\times 10^{-2}$ we find $|\beta|\lesssim 1$.

As the Fig. \ref{fig:parameter-space} shows, typical values of parameters are ${\cal G}\sim 10^{-2}$, $\mathbb{H}\sim H\sim \mathrm{few}\times 10^{-5} M_{Pl}$, $\omega\sim 10^{-3}$ and $f\sim \mathrm{few}\times 10^{-3} M_{Pl}$. For these values, indeed the one-loop and tree level contributions to power spectrum of curvature perturbations \eqref{PR-our-result} are of the same order and indeed the one-loop result can dominate over the tree level by a factor of two to four.

\subsection{A new shape for non-Gaussianity}\label{sec:NGy}

The three-point function \eqref{bispectrum-oneloop} is related to the bispectrum in the CMB temperature fluctuations which is usually written in terms of  the $f_{NL}$ factor and a shape function \cite{Akrami:2019izv}, 
\be
B_{\varphi}(\k_1,\k_2,\k_3):=\left(\frac{5}{3}\frac{\dot\phi}{H}\right)^3\dfrac{6 A_s^2 f^{\mathrm{res.}}_{NL}}{(k_1\,k_2\,k_3)^2}S(k_1,k_2,k_3).
\ee
Here the prefactor $\frac{5}{3}\frac{\dot\phi}{H}$ is added to relate the amplitude of  the gauge invariant Bardeen potential in matter dominated era to the inflaton fluctuations $\vphi$ and $A_s$ is the dimensionless power spectrum which in our case is
\be
A_s:=\frac{9}{25} P_{\cal R}=\frac{9}{25} \left(\frac{H}{2\pi}\right)^2\left(\frac{H}{\dot\phi}\right)^2\left[1+ \dfrac{\pi^3}{24}\left( \dfrac{\mathbb H}{H}\right)^2\right]\simeq \frac{9\pi^2}{800}\frac{{\cal G}^2}{\alpha^2},
\ee
and finally, 
\be
S(k_1,k_2,k_3):= \dfrac{\sin \Phi(2k_1,\alpha)+\sin \Phi(2k_2,\alpha)+\sin \Phi(2k_3,\alpha)}{3},
\ee
is the `shape function.'  This shape function is  superficially similar to that of resonance and axion monodromy models \cite{Akrami:2019izv,Chen:2010bka}. However, it must be noted that it has a completely different momentum dependence. A simple algebra then yields 
\be
f^{\mathrm{res.}}_{NL}=640 \sqrt{2\pi} \dfrac{\ln \alpha}{{\cal G}\,\alpha^{3/2}}.
\ee
For example, for typical parameters $\alpha=100$ and ${\cal G}=0.03$ we get $f_{NL}\simeq 250$. This shows that this model predicts a not so small non-Gaussianity with a shape which is somewhat supported by the data.


There is  evidence, though of low significance, for the resonant non-Gaussianity with  modulation frequency $\alpha \in [ 0,50]$ \cite{Akrami:2019izv}, see also \cite{Munchmeyer:2019wlh}. However, observations of the CMB temperature fluctuations of are unable to constrain \emph{amplitude} of this type of non-Gaussianity \cite{Akrami:2019izv}. It has been recently shown that observations of galaxy clustering by future surveys will not improve CMB bounds on the amplitude of resonant non-Gaussianity   \cite{Cabass:2018roz}. Nonetheless, from theoretical perspective, assuming non-Gaussian corrections to the scalar perturbation to be smaller than the Gaussian part, for $\expect{\cal R}_c^2 \sim 10^{-9}$, we get $f^{\mathrm{res.}}_{NL} \lesssim 10^4$.  Nevertheless, upcoming surveys of HI intensity map like SKA can potentially reduce the uncertainty of amplitude of resonant non-Gausssianity to $\sigma(f^{\mathrm{res.}}_{NL}) \sim 10$ through studying the so-called scale-dependent bias \cite{Bacon:2018dui}.

\section{Summary and concluding remarks}\label{sec:discussion}

In this work we studied  a specific slow-roll inflationary setup in which background inflationary trajectory is that of a usual, e.g. single field chaotic, inflationary model, while its cosmic perturbation theory is totally different. In our setup besides the inflaton $\phi$ there is another light scalar field $\chi$  which is coupled to the inflaton which we assumed not to have a self-interacting potential. The potential for the $\phi$  field besides the term which drives slow-roll inflation, has a cosine part. This part oscillates with frequency $\omega=\alpha H$ during inflation and has a discrete shift symmetry $\phi\to \phi+2\pi f$, where $f\sim \omega$. This oscillatory part contributes to the energy budget during inflation at percent level and hence does not change the inflationary trajectory. There is also another oscillatory mass term for the $\chi$ field which exhibits the same discrete shift symmetry.
The effects of such a cosine potential term for the inflaton has been widely studied in the context of (generalized) monodromy inflation models and in this work we focused on the effects of the $\chi$ sector. Given that such light scalar fields with cosine modulated potentials are ubiquitous in multi-axion models.

The $\chi$ sector, by construction, does not contribute to classical inflationary trajectory. Our model may hence be viewed as a typical two field model where the $\chi$ field is the isocurvature mode. As in usual two-field models the tree level power spectrum of the isocurvature modes is small compared to that of the inflaton. However,  the oscillatory coupling between $\chi$ and $\phi$ sectors leads to novel, observable loop effects, which we studied in details. We established that the spectator (isocurvature) $\chi$ modes, while running in the loops, can  amplify the power spectrum of curvature perturbations. This latter leads to a suppression of tensor-to-scalar ratio $r$. They can also change the spectral tilt $n_s$. These two are basically the two main observables of inflationary models restricted by Planck data. The suppression of $r$ in our model allows for reconciliation of chaotic inflationary models with the Planck data \cite{Resonant-Monodromy}. 

In our model we have two dimensionful parameters $\omega, f$ and a dimensionless coupling ${\cal G}$. Typical theoretically and observationally  motivated range for these parameters are $\omega\lesssim f\sim 10^{-3} M_{Pl}$ and ${\cal G}\sim 10^{-2}$. 
These values for $\omega, f$ which are close to (but slightly less than) the GUT scale, make our power-suppression scenario to be naturally realizable within the GUT models and the axion fields there.

 We also analyzed in details the three-point function of scalar perturbations and showed that the one-loop result dominates over the higher loop corrections. We found that our model has a novel shape of non-Gaussianity, similar to, but different than, the resonant non-Gaussianity \cite{Cabass:2018roz}. The $f_{NL}$ factor for our model is large enough  which may be observable in future \cite{Bacon:2018dui}. 

We discussed that resonant production of $\chi$ particles during inflation is negligible. However, as the inflation ends and inflaton roles toward the minimum of its potential, the narrow band $\chi$ production also enhances yielding a more efficient (p)reheating scenario. It would be interesting to study its possible observable effects on primordial gravity wave production.

Finally, in this work we developed a lot of techniques and tools for the in-in QFT computations. These could be useful  in other problems whenever in-in formulation is used and/or when we have a time-dependent oscillatory coupling in condensed matter problems.


\section*{Acknowledgement}
We are grateful to Yashar Akrami and Mehrdad Mirbabayi  for fruitful discussions and Eva Silverstein, Xingan Chen and Sadra Jazayeri for comments. MMShJ's work is supported in part by INSF grant No.~950124 and Saramadan grant ISEF/M/98204. We would like to thank the hospitality of ICTP HECAP where  this work carried out. AAA. is partially supported by School of Physics of Institute of Research in Fundamental Sciences (IPM).

\appendix
\section{Details of  one-loop two-point function the time integrals}\label{appen:two-pint-one-loop}

The one-loop two-point function \eqref{2pf} is sum of the following two integrals
\be\label{P11-P2}
P_{11}(k)=\Pi (k)\  \int \dfrac{d^3\p}{pk|-\p+\k|} \ {\mathbf{I}_{11}(p,k)},\qquad
P_{2}(k)= \Pi(k)\  \int \dfrac{d^3\p}{pk|-\p+\k|} \ {\mathbf{I}_{2}(p,k)},
\ee
where 
\be 
\Pi(k)\equiv \frac{{\cal G}^2 f^2 P_{\phi}(k)}{2},
\ee
\be\begin{split}
{\mathbf{I}_{11}(p,k)}&=\int^0_{-\infty}  \frac{d\tau_1}{H\tau_1} \sin \omega t_1 (\cos k\tau_1-\frac{\sin k\tau_1}{k\tau_1})
\int^{\tau_1}_{-\infty} \frac{d\tau_2}{H\tau_2} \sin \omega t_2 (\cos k\tau_2-\frac{\sin k\tau_2}{k\tau_2}) \cos\ell(\tau_1-\tau_2),\\
{\mathbf{I}_{2}(p,k)}&=\int^0_{-\infty}  \frac{d\tau_1}{H\tau_1} \sin \omega t_1 (\cos k\tau_1-\frac{\sin k\tau_1}{k\tau_1})
\int^{\tau_1}_{-\infty} \frac{d\tau_2}{H\tau_2} \sin \omega t_2 (\sin k\tau_2+\frac{\cos k\tau_2}{k\tau_2}) \sin\ell(\tau_1-\tau_2),
\end{split}\ee
with $\ell:=p+|\k-\p|$.

The integrals over $\tau$ cannot be computed in a closed form. Recalling that the integrand has an oscillatory behavior one may use stationary phase approximation to evaluate the integrals. In particular, recalling that $t=\frac{1}{H}\ln(-H\tau)$, we consider the following two basic integrals:
\be\label{Int-1}\int^{\tau} \dfrac{d \tau'}{\tau'} e^{i\omega t'}\, e^{- i K_t\tau'} = \int^{\tau} \dfrac{d \tau'}{\tau'  }  e^{- i K_t \tau + i\frac{\omega}{H} \ln(\tau/\tau_{\ast})}\simeq 0
\ee
for $\tau<0$ and $\omega, K_t>0$. This is due to the fact that the stationary phase point happens at $\tau_0=\frac{\omega}{K_t H}>0$ which is not in the range of the integral for $\tau<0$. 
The second integral is 
\be\label{int-2}
\int^{\tau} \dfrac{d \tau'}{\tau'} e^{-i\omega t'}\, e^{- i K_t \tau'} = \int^{\tau} \dfrac{d \tau'}{\tau'  }  e^{- i K_t\tau - i\frac{\omega}{H} \ln(\tau/\tau_{\ast})}
\simeq \sqrt{\dfrac{2\pi }{\,\alpha}} e^{i\left(\alpha-\alpha \ln \alpha +\alpha \ln (-K_t\tau_{\ast})-\frac{\pi}{4}\right)}\theta(\tau+\tau_s),
\ee
where $\alpha=\omega/H$,
\be\label{eta-s}
\tau_s\approx\tau_0(1+\frac{1}{\sqrt{\alpha}}),\qquad \tau_0=\frac{\alpha}{K_t},
\ee
$\tau_{\ast}$ denotes some arbitrary time, $\theta(\tau+\tau_s)$ is the step function and $\tau_0$ is where the saddle point is and $\frac{\tau_0}{\sqrt{\alpha}}$ is the width of the Gaussian around the saddle. We note that the saddle point (or steady phase) approximation gives a good estimate of the integral and the step function should not be viewed as a sharp cutting function for the integral. Therefore, 
\be\label{saddle-point:1}
\int^{\tau} \dfrac{d \tau'}{\tau'} \sin\omega t'\, e^{\pm i K_t \tau'}  
\approx \mp i\sqrt{\dfrac{\pi }{2\,\alpha}} e^{\pm i\Phi(K_t,\alpha)}\ \theta(\tau+\tau_s),
\ee
and analogously, 
\be\label{saddle-point:2}
\int^{\tau} \dfrac{d \tau'}{K_t\tau'^2} \sin\omega t'\, e^{\pm i K_t \tau'}  
\approx \pm i \sqrt{\dfrac{\pi }{2\,\alpha^3}} e^{\pm i\Phi(K_t,\alpha)}\ \theta(\tau+\tau_s),
\ee
where
\be\label{Phi}
\Phi(K_t, \alpha)=\alpha \ln \alpha -\alpha-\alpha \ln (-K_t\tau_{\ast})+\frac{\pi}{4}.
\ee
Note that in our computations $K_t$ is the \emph{net energy transfer} at the vertex, that is, $K_t$ is the difference between the zero's component of the momentum four-vector of fields appearing at a vertex. Recall that on an expanding background, the uncertainty principle allows for violation of the energy conservation of the order $H$. In our case, the oscillating coupling constant $g_{3}(t)$ is a more important source for  energy non-conservation. Here, each vertex can transfer four momentum $p^{\mu}=(\omega,0,0,0)$ from/to the background and $\omega=\alpha H$. In fact RHS of \eqref{saddle-point:2} is what replaces the energy conservation delta-function.

Therefore, 
\be\begin{split}
{\mathbf{I}^{(2)}_{11}(\ell,k)}&:=\int^{\tau_1}_{-\infty} \frac{d\tau_2}{H\tau_2} \sin \omega t_2 (\cos k\tau_2-\frac{\sin k\tau_2}{k\tau_2}) \cos\ell(\tau_1-\tau_2)\cr
&=-\sqrt{\dfrac{\pi }{2\,\alpha}}\frac{1}{2H}[A\cos \ell\tau_1+ B\sin \ell\tau_1]\theta(\tau+\tau_s)
\end{split}
\ee
where $\ell=p+|\k-\p|$ and
\be\begin{split}
A&=-\sin\Phi_+- \sin\Phi_-+\frac{1}{\alpha}(\frac{\ell}{k}+1)\cos\Phi_+- \frac{1}{\alpha}(\frac{\ell}{k}-1)\cos\Phi_-,\quad \Phi_-\equiv \Phi(\ell-k,\alpha),\cr
B&=\cos\Phi_++ \cos\Phi_-+\frac{1}{\alpha}(\frac{\ell}{k}+1)\sin\Phi_+- \frac{1}{\alpha}(\frac{\ell}{k}-1)\sin\Phi_-,\quad \Phi_+\equiv \Phi(\ell+k,\alpha),
\end{split}
\ee
and
\be\begin{split}
{\mathbf{I}^{(2)}_{2}(\ell,k)}&:=\int^{\tau_1}_{-\infty} \frac{d\tau_2}{H\tau_2} \sin \omega t_2 (\sin k\tau_2+\frac{\cos k\tau_2}{k\tau_2}) \sin(p+|\k-\p|)(\tau_1-\tau_2)
\cr
&=\sqrt{\dfrac{\pi }{2\,\alpha}}\frac{1}{2H}[C\cos\ell\tau_1+ D\sin\ell\tau_1]\theta(\tau+\tau_s)
\end{split}
\ee
where 
\be\begin{split}
C&=\sin\Phi_+- \sin\Phi_--\frac{1}{\alpha}(\frac{\ell}{k}+1)\cos\Phi_+- \frac{1}{\alpha}(\frac{\ell}{k}-1)\cos\Phi_-,\cr
D&=-\cos\Phi_++\cos\Phi_--\frac{1}{\alpha}(\frac{\ell}{k}+1)\sin\Phi_+- \frac{1}{\alpha}(\frac{\ell}{k}-1)\sin\Phi_-,   
\end{split}
\ee
\paragraph{Computation of ${\mathbf{I}_{11}(p,k)}$.}

\be\label{I-11}
\begin{split}
{\mathbf{I}_{11}(p,k)}&=\int^0_{-\infty}  \frac{d\tau_1}{H\tau_1} \sin \omega t_1 (\cos k\tau_1-\frac{\sin k\tau_1}{k\tau_1})\times \mathbf{I}^{(2)}_{11}(\ell,k)= \frac{\pi}{8H^2\alpha} (A^2+B^2)\cr
&= \frac{\pi}{2H^2\alpha} \left[(\cos\Theta+\frac{\ell}{k\alpha}\sin\Theta)^2+\frac{1}{\alpha^2}\cos^2\Theta\right],
\end{split}
\ee
where $2\Theta=\Phi_+-\Phi_-=-\alpha\ln(\frac{\ell+k}{\ell-k})$.
\paragraph{Computation of ${\mathbf{I}_{2}(p,k)}$.} 
\be\label{I-2}
\begin{split}
{\mathbf{I}_{2}(p,k)}&=\int^0_{-\infty}  \frac{d\tau_1}{H\tau_1} \sin \omega t_1 (\cos k\tau_1-\frac{\sin k\tau_1}{k\tau_1})\times \mathbf{I}^{(2)}_{2}(\ell,k)=-\frac{\pi}{8H^2\alpha} (AC+BD)\cr
&= \frac{\pi}{2H^2\alpha^2}\frac{\ell}{k\alpha}.
\end{split}
\ee
Both of \eqref{I-11} and \eqref{I-2} are valid provided that  $\theta(\tau+\tau_s)$ factors is nonzero. To verify this let, 
\[\kappa_\pm=\ell\pm k=p+|\p-\k|\pm k.\]
The $\kappa_\pm$ terms contribute to the integrals at the saddle points $\kappa_\pm\tau_\pm=-{\alpha}$. Of course the integrals are nonzero if the saddle points $\tau^\pm$ are within the $[-\tau_{1s}, 0]$ range. Recalling that $\kappa_+=\kappa$, $-\tau_+<\tau_{1s}$ and hence the $\kappa_+$ phase always contributes to the integral. For the $\tau_-$, however, we need to make sure that 
\be\kappa_-\gtrsim \frac{\kappa_+}{\sqrt{\alpha}+1}, \quad \text{or}\qquad  \frac{k}{\ell}\lesssim \frac{\sqrt{\alpha}}{2+\sqrt{\alpha}}.
\ee
The above yields $p\gtrsim k(1+\frac{1}{\sqrt{\alpha}})$.

\section{Details of calculations of one-loop three-point function}
\label{Appen:bispectrum-full}
Here we present a detailed calculation of the one-loop three-point function of scalar fluctuations, i.e. \eqref{3-pt-one-loop-1}. At one-loop level there are some different three-point functions to compute:
\be
B_{ijk}(\k_1,\k_2,\k_3):=\expect{\vphi_i(\k_1)\vphi_j(\k_2) \vphi_k(\k_3)}
\ee
where $i,j,k$ are taking values among $0,1,2,3$ such that $i+j+k=3$. 

Associated Feynman diagrams are depicted in Fig \ref{Bispectrum:Feyn-Dia-one-loop}. Note that, in general, a single diagram can be representative of a handful of correlation functions or conversely a single term in the perturbative expansion produce few diagrams.
One-loop diagram (a) corresponds to $i=j=k=1$ case. The diagram (b) corresponds to $i\neq j\neq k$, which means one of them should be $0$, one should be $1$ and the last one should be $2$. Therefore, there are six such choices. Diagrams (c), (d) correspond to the cases where among $i,j,k$ two are $0$ and the last one is $3$. There are hence three such cases. 
This calculation is rather lengthy and cumbersome and we have confirmed our results using a Mathematica code. However, here we present first few steps of the analysis. It is useful to define the following shorthand notation, 
\begin{equation}
\hspace*{-5mm}
\iiint_{\tau} \mathbf{X} \equiv P_{\phi}^{1/2}(k_1)P_{\phi}^{1/2}(k_2)P_{\phi}^{1/2}(k_3)\int \dfrac{d\tau_1}{H^4\tau^4_1} \dfrac{d\tau_2}{H^4\tau^4_2} \dfrac{d\tau_3}{H^4\tau^4_3}~g_{3}(\tau_1)g_{3}(\tau_2)g_{3}(\tau_3)   \cdot \mathbf{X}   
\end{equation}
for any function $\mathbf{X}$. 

Let us start  with the simplest case, namely diagram (a) and $B_{111}$. Using \eqref{varphi-1} we have 
\begin{align}
B_{111}(k_1,k_2,k_3)&\equiv \expect{\vphi_{1}(\k_1)\vphi_{1}(\k_2)\vphi_{1}(\k_3)}
\\
\nonumber
&=  8~ \iiint_{\tau} \mathrm{Re} \vphi_{\k_1}(\tau_1) \mathrm{Re} \vphi_{\k_2}(\tau_2) \mathrm{Re} \vphi_{\k_3}(\tau_3)~ \big\langle (\hat{\chi}^2)_{\k_1}(\tau_1) (\hat{\chi}^2)_{\k_2}(\tau_2) (\hat{\chi}^2)_{\k_3}(\tau_3) \big \rangle 
\\
\nonumber
&=8\times2 ~ \int d^3\p \iiint_{\tau} \mathrm{Re} \vphi_{\k_1}(\tau_1) \mathrm{Re} \vphi_{\k_2}(\tau_2) \mathrm{Re} \vphi_{\k_3}(\tau_3)\times
\\
\nonumber
&~~~~~~~~~~~~~~~~~~~~~~~~~~\bigg[\,\chi_{\p}(\tau_1) \chi_{-\p+\k_1}(\tau_1)\,\chi_{\p+\k_2}(\tau_2)\chi^{\ast}_{-\p}(\tau_2) \, \chi^{\ast}_{\p-\k_1}(\tau_3)\chi^{\ast}_{-\p-\k_2}(\tau_3)
\\
\nonumber
&~~~~~~~~~~~~~~~~~~~~~~~~~~+\chi_{\p}(\tau_1) \chi_{-\p+\k_1}(\tau_1)\,\chi_{-\p-\k_3}(\tau_2)\chi^{\ast}_{\p-\k_1}(\tau_2) \, \chi^{\ast}_{-\p}(\tau_3)\chi^{\ast}_{\p+\k_3}(\tau_3) \bigg]
\end{align}
As for the diagram (b) and the corresponding six  choices, using \eqref{varphi-2} and \eqref{varphi-1}, we have
\begin{align}
\hspace*{-.3 cm}
B_{210}(k_1,k_2,k_3)&\equiv \expect{\vphi_{2}(\k_1)\vphi_{1}(\k_2)\vphi_{0}(\k_3)}
\\ \nonumber &= -8\,i\int d^3\p\iiint_{\tau}\mathrm{Re} \vphi_{\k_1}(\tau_1) \mathrm{Re} \vphi_{\k_2}(\tau_3) \vphi_{\k_3}(\tau_2) G_{\chi}(\tau_1,\tau_2;\p)\times \\ \nonumber &\qquad\qquad\qquad \times\big\langle \{\hat{\chi}_{-\p+\k_1}(\tau_1), \hat{\chi}_{\p+\k_2}(\tau_2)\} \left(\hat{\chi}^2\right)_{\k_3}(\tau_3) \big\rangle
\\ \nonumber
&= 32 \int d^3\p \iiint_{\tau} \mathrm{Re} \vphi_{\k_1}(\tau_1) \mathrm{Re} \vphi_{\k_2}(\tau_3) \vphi_{\k_3}(\tau_2) \left(\chi_{\p}(\tau_1)\chi^{\ast}_{\p}(\tau_2)-\chi^{\ast}_{\p}(\tau_1)\chi_{\p}(\tau_2)\right) \times
\\
\nonumber
&~~~~~~~~~~~~~~~~~~~~~~\big [\chi_{-\p+\k_1}(\tau_1)\,\chi_{\p+\k_3}(\tau_2) \chi^{\ast}_{\p-\k_1}(\tau_3)\chi^{\ast}_{-\p-\k_3}(\tau_3)\big]
\end{align}
\begin{align}
\hspace*{-.3 cm}
B_{201}(k_1,k_2,k_3)&\equiv \expect{\vphi_{2}(\k_1)\vphi_{0}(\k_2)\vphi_{1}(\k_3)}
\\ \nonumber &= -8\, i\int d^3\p\iiint_{\tau}\mathrm{Re} \vphi_{\k_1}(\tau_1) \mathrm{Re} \vphi_{\k_3}(\tau_3) \vphi_{\k_2}(\tau_2) G_{\chi}(\tau_1,\tau_2;\p) \times \\ \nonumber &\qquad\qquad\qquad \times \big\langle \{\hat{\chi}_{-\p+\k_1}(\tau_1), \hat{\chi}_{\p+\k_2}(\tau_2)\} \left(\hat{\chi}^2\right)_{\k_3}(\tau_3) \big\rangle
\\ \nonumber
&= 32\int d^3\p\iiint_{\tau} \mathrm{Re} \vphi_{\k_1}(\tau_1) \mathrm{Re} \vphi_{\k_3}(\tau_3) \vphi_{\k_2}(\tau_2) \left(\chi_{\p}(\tau_1)\chi^{\ast}_{\p}(\tau_2)-\chi^{\ast}_{\p}(\tau_1)\chi_{\p}(\tau_2)\right) \times
\\
\nonumber
&~~~~~~~~~~~~~~~~~~~~~~\big [\chi_{-\p+\k_1}(\tau_1)\,\chi_{\p+\k_2}(\tau_2) \chi^{\ast}_{\p-\k_1}(\tau_3)\chi^{\ast}_{-\p-\k_2}(\tau_3)\big],
\end{align}
\begin{align}
\hspace*{-.3 cm}
B_{120}(k_1,k_2,k_3)&\equiv \expect{\vphi_{1}(\k_1)\vphi_{2}(\k_2)\vphi_{0}(\k_3)}
\\ \nonumber &= -8 i\int d^3\p\iiint_{\tau}\mathrm{Re} \vphi_{\k_1}(\tau_3) \mathrm{Re} \vphi_{\k_2}(\tau_1) \vphi_{\k_3}(\tau_2) G_{\chi}(\tau_1,\tau_2;\p) \times \\ \nonumber &\qquad\qquad\qquad \times\big\langle \left(\hat{\chi}^2\right)_{\k_1}(\tau_3) \{\hat{\chi}_{-\p+\k_2}(\tau_1), \hat{\chi}_{\p+\k_3}(\tau_2)\}  \big\rangle
\\ \nonumber
&= 32\int d^3\p\iiint_{\tau} \mathrm{Re} \vphi_{\k_1}(\tau_3) \mathrm{Re} \vphi_{\k_2}(\tau_1) \vphi_{\k_3}(\tau_2) \left(\chi_{\p}(\tau_1)\chi^{\ast}_{\p}(\tau_2)-\chi^{\ast}_{\p}(\tau_1)\chi_{\p}(\tau_2)\right) \times
\\
\nonumber
&~~~~~~~~~~~~~~~~~~~~~~\big [\chi^{\ast}_{-\p+\k_2}(\tau_1)\,\chi^{\ast}_{\p+\k_3}(\tau_2) \chi_{\p-\k_2}(\tau_3)\chi_{-\p-\k_3}(\tau_3)\big],
\end{align}
and
\begin{align}
    B_{102}(k_1,k_2,k_3) &= B^{\ast}_{201}(k_3,k_2,k_1),
    \\
    B_{021}(k_1,k_2,k_3) &= B^{\ast}_{120}(k_3,k_2,k_1),
    \\
    B_{012}(k_1,k_2,k_3) &= B^{\ast}_{210}(k_3,k_2,k_1).
\end{align}

We next  compute the contributions of diagrams (c) and (d).  These contributions involve $\vphi_3$ given in \eqref{varphi-3}. The first term of $\vphi_3$ in \eqref{varphi-3} contributes \emph{disconnected} diagrams in the three-point function and can be omitted. Then, we obtain 
\begin{align}
B_{300c}(k_1,k_2,k_3)&\equiv \expect{\vphi_{3}(\k_1)\vphi_{0}(\k_2)\vphi_{0}(\k_3)}_c
\\
\nonumber
&= - 2\times 4 \int d^3\p \int d^3 \k' P_{\vphi}(k_2)^{-1/2}P_{\vphi}(k_3)^{-1/2}\iiint_{\tau} \mathrm{Re} \vphi_{\k_1}(\tau_1) G_{\chi}(\tau_1,\tau_2;\p) \times 
\\
\nonumber
&~~~~~~~~~~~~~~~~G_{\chi}(\tau_2,\tau_3;\p+\k')~~\big\langle \{\hat{\chi}_{-\p+\k_1}(\tau_1), \{ \hat{\vphi}_{-\k'}(\tau_2), (\hat{\vphi}(\tau_3)\hat{\chi}(\tau_3))_{\p+\k'} \}\} \big\rangle
\\
\nonumber
&= -2\times 4\int d^3\p \iiint_{\tau} \mathrm{Re}\vphi_{\k_1}(\tau_1) \left(\chi_{-\p+\k_1}(\tau_1)\chi_{\p-\k_1}^{\ast}(\tau_3)+ \chi_{-\p+\k_1}^{\ast}(\tau_1)\chi_{\p-\k_1}(\tau_3) \right)\times
\\ 
\nonumber 
&~~
G_{\chi}(\tau_1,\tau_2;\p)\bigg[G_{\chi}(\tau_1,\tau_2;\p+\k_2) \vphi_{\k_3}(\tau_3)\vphi_{\k_2}(\tau_2)+G_{\chi}(\tau_1,\tau_2;\p+\k_3) \vphi_{\k_2}(\tau_3)\vphi_{\k_3}(\tau_2)
\bigg],
\end{align}
\begin{align}
B_{030c}(k_1,k_2,k_3)&\equiv \expect{\vphi_{0}(\k_1)\vphi_{3}(\k_2)\vphi_{0}(\k_3)}_c
\\
\nonumber
&= - 2\times 4 \int d^3\p \int d^3 \k' P_{\vphi}(k_1)^{-1/2}P_{\vphi}(k_3)^{-1/2}\iiint_{\tau} \mathrm{Re} \vphi_{\k_2}(\tau_1) G_{\chi}(\tau_1,\tau_2;\p) \times 
\\
\nonumber
&~~~~~~~~~~~~~~G_{\chi}(\tau_2,\tau_3;\p+\k')~~\big\langle \{\hat{\chi}_{-\p+\k_2}(\tau_1), \{ \hat{\vphi}_{-\k'}(\tau_2), (\hat{\vphi}(\tau_3)\hat{\chi}(\tau_3))_{\p+\k'} \}\} \big\rangle
\\
\nonumber
&= -2\times 4\int d^3\p \iiint_{\tau} \mathrm{Re}\vphi_{\k_2}(\tau_1) \left(\chi_{-\p+\k_2}(\tau_1)\chi_{\p-\k_2}^{\ast}(\tau_3)+ \chi_{-\p+\k_2}^{\ast}(\tau_1)\chi_{\p-\k_2}(\tau_3) \right)\times
\\ 
\nonumber 
&~~ 
G_{\chi}(\tau_1,\tau_2;\p)\bigg[G_{\chi}(\tau_1,\tau_2;\p+\k_2) \vphi_{\k_3}(\tau_3)\vphi_{\k_1}^{\ast}(\tau_2)+G_{\chi}(\tau_1,\tau_2;\p+\k_3) \vphi_{\k_1}^{\ast}(\tau_3)\vphi_{\k_3}(\tau_2)
\bigg],\nonumber
\end{align}
and $B_{003c}(k_1,k_2,k_3)$ can be simply found, as 
\begin{align}
B_{003c}(k_1,k_2,k_3)= B^{\ast}_{300c}(k_3,k_2,k_1).  
\end{align}
Contribution of the  diagram (d) can be computed in a similar way,
\begin{align}
B_{300d}(k_1,k_2,k_3)&\equiv \expect{\vphi_{2}(\k_1)\vphi_{0}(\k_2)\vphi_{1}(\k_3)}_c
\\
\nonumber
&=  -2\times 4\int d^3\p \int d^3 \k'd^3 \k'' P_{\vphi}(k_2)^{-1/2}P_{\vphi}(k_3)^{-1/2}\iiint_{\tau} \mathrm{Re} \vphi_{\k_1}(\tau_1) G_{\chi}(\tau_1,\tau_2;\p)\times
\\
\nonumber
&~~~~~~~~~~~~~~
 G_{\chi}(\tau_1,\tau_3;-\p+\k_1)\times 
\big \langle
\{\hat{\chi}_{\p+\k'}(\tau_2)
\{\hat{\chi}_{-\p-\k'}(\tau_3)
\big \rangle
\big\langle  \{ \hat{\vphi}_{-\k'}(\tau_2), \hat{\vphi}_{-\k''}(\tau_3) \}\} \big\rangle
\\
\nonumber
&= 2\times4\int d^3\p \iiint_{\tau} \mathrm{Re}\vphi_{\k_1}(\tau_1) 
G_{\chi}(\tau_1,\tau_2;\p)G_{\chi}(\tau_1,\tau_2;-\p+\k_1)\times 
\\
\nonumber
&~~~~~~~~~~~~~~~\bigg \{\left(\chi_{\p+\k_2}(\tau_2)\chi_{-\p-\k_2}^{\ast}(\tau_3)+ \chi_{-\p-\k_2}^{\ast}(\tau_2)\chi_{\p+\k_2}(\tau_3) \right) \vphi_{\k_3}(\tau_3)\vphi_{\k_2}(\tau_2)+
\\
\nonumber
&~~~~~~~~~~~~~~\left(\chi_{\p+\k_3}(\tau_2)\chi_{-\p-\k_3}^{\ast}(\tau_3)+ \chi_{-\p-\k_3}^{\ast}(\tau_2)\chi_{\p+\k_3}(\tau_3) \right) \vphi_{\k_3}(\tau_2)\vphi_{\k_2}(\tau_3)\bigg \},
\end{align}
\begin{align}
B_{030d}(k_1,k_2,k_3)&\equiv \expect{\vphi_{2}(\k_1)\vphi_{0}(\k_2)\vphi_{1}(\k_3)}_c
\\
\nonumber
&=  -2\times 4\int d^3\p \int d^3 \k'd^3 \k'' P_{\vphi}(k_1)^{-1/2}P_{\vphi}(k_3)^{-1/2}\iiint_{\tau} \mathrm{Re} \vphi_{\k_2}(\tau_1) G_{\chi}(\tau_1,\tau_2;\p)\times
\\
\nonumber
&~~~~~~~~~~~~~~
 G_{\chi}(\tau_1,\tau_3;-\p+\k_2)\times 
\big \langle
\{\hat{\chi}_{\p+\k'}(\tau_2)
\{\hat{\chi}_{-\p-\k'}(\tau_3)
\big \rangle
\big\langle  \{ \hat{\vphi}_{-\k'}(\tau_2), \hat{\vphi}_{-\k''}(\tau_3) \}\} \big\rangle
\\
\nonumber
&= 2\times4\int d^3\p \iiint_{\tau} \mathrm{Re}\vphi_{\k_2}(\tau_1) 
G_{\chi}(\tau_1,\tau_2;\p)G_{\chi}(\tau_1,\tau_2;-\p+\k_2)\times 
\\
\nonumber
&~~~~~~~~~~~~~~~~\bigg \{\left(\chi_{\p+\k_1}(\tau_2)\chi_{-\p-\k_1}^{\ast}(\tau_3)+ \chi_{-\p-\k_1}^{\ast}(\tau_2)\chi_{\p+\k_1}(\tau_3) \right) \vphi_{\k_3}(\tau_3)\vphi^{\ast}_{\k_1}(\tau_2)+
\\
\nonumber
&~~~~~~~~~~~~~~~\left(\chi_{\p+\k_3}(\tau_2)\chi_{-\p-\k_3}^{\ast}(\tau_3)+ \chi_{-\p-\k_3}^{\ast}(\tau_2)\chi_{\p+\k_3}(\tau_3) \right) \vphi_{\k_3}(\tau_2)\vphi^{\ast}_{\k_1}(\tau_3)\bigg \},
\end{align}
and 
\begin{align}
B_{003d}(k_1,k_2,k_3)= B^{\ast}_{300d}(k_3,k_2,k_1).  
\end{align}

The above involve three time integrals and a loop momentum integral. The time integrals are receive dominant contribution around the resonance time $\tau_s$ where $k\tau_s\gtrsim 1$, where $k$ is a typical value of the external momenta $k_i$. The propagators associated with the outgoing legs, however, then yield to UV cutoff on the loop momentum of order $p_{_{UV}}\sim \alpha k$. In what follows we show the last final results for each of the diagrams.

\paragraph{Diagram (b).} 
\begin{align}
\label{B210b:I}
B_{210}&+B_{201}+B_{120}+B_{102}+B_{021}+B_{012}= i\frac{  f^3 {\cal G}^3}{32 \sqrt{2}}\dfrac{1}{ k_1^2 k_2^2 k_3^2} \left(\frac{\pi}{\alpha }\right)^{3/2} \int^{\alpha k} \dfrac{d p}{p}\int d \Omega_{\p}\times
\\
\nonumber
 &\bigg[ +e^{i \Phi (k_{1-},\alpha)}-e^{i \Phi (k_{1+},\alpha)}
+ e^{-i \Phi (k_{2-},\alpha)}- e^{i \Phi (k_{2+},\alpha)}
-e^{-i \Phi (k_{3-},\alpha)}+ e^{-i \Phi (k_{3+},\alpha)}\bigg]. 
\end{align}
One may readily observe that the first and last couple of terms cancel out, as expected from the parity symmetry, $\p \rightarrow -\p$.

\paragraph{Diagram (c).} 
\begin{align}
\nonumber
B_{300(c)}&+B_{030(c)}+B_{003(c)}= i\frac{  f^3 {\cal G}^3}{256 \sqrt{2}}\dfrac{1}{ k_1^2 k_2^2 k_3^2} \left(\frac{\pi}{\alpha }\right)^{3/2}\int^{\alpha k}\dfrac{d p}{p} \int d\Omega_{\p}
\\
&  \bigg[ e^{i \Phi (k_{1-},\alpha)}-e^{i \Phi (k_{1+},\alpha)}-3\,e^{-i \Phi (k_{2-},\alpha)}+3\,e^{i \Phi (k_{2+},\alpha)}
-e^{-i \Phi (k_{3-},\alpha)}+e^{-i \Phi (k_{3+},\alpha)}+\frac12 {\cal I}_{3\Phi(c)}\bigg ].
\label{B300c:I}
\end{align}
\paragraph{Diagram (d).} 
\begin{align}
\nonumber
&B_{300(d)}+B_{030(d)}+B_{003(d)}= i\frac{  f^3 {\cal G}^3}{256 \sqrt{2}}\dfrac{1}{ k_1^2 k_2^2 k_3^2} \left(\frac{\pi}{\alpha }\right)^{3/2} \int^{\alpha k}\dfrac{d p}{p} \int d\Omega_{\p} 
\\
&\bigg[- e^{i \Phi (k_{1-},\alpha)}+e^{i \Phi (k_{1+},\alpha)}+5\,e^{-i \Phi (k_{2-},\alpha)}-5\,e^{i \Phi (k_{2+},\alpha)}
+e^{-i \Phi (k_{3-},\alpha)}-e^{-i \Phi (k_{3+},\alpha)}+ {\cal I}_{3\Phi(d)}\bigg ].
\label{B300d:I}
\end{align}
In the above ${\cal I}_{3\Phi(c)}$ are ${\cal I}_{3\Phi(d)}$ are conditional expressions containing exponents of  sums of three different phase functions
$\exp [{\pm i \Phi (k_{1 \pm}) \pm i \Phi (k_{2 \pm})\pm i \Phi (k_{3 \pm})}]$. Theses terms belong to the expressions that have small energy transfer of the background. We  show below that using the stationary phase approximation, the momentum integral of these terms are identically zero.
A direct computation leads to
\begin{align}
\nonumber
&{\cal I}_{3\Phi(c)}=
\\
\nonumber
&- e^{-i (\Phi (k_{1-})+\Phi (k_{2-})+\Phi (k_{3-}))} \theta (k_{1-}-k_{2-},k_{2-}-k_{3-})+ e^{+i (\Phi (k_{1-})+\Phi (k_{2-})+\Phi (k_{3-}))} \theta (k_{1-}-k_{2-},k_{3-}-k_{1-})
\\
\nonumber
&+ e^{+i (\Phi (k_{1-})+\Phi (k_{2-})+\Phi (k_{3-}))} \theta (k_{2-}-k_{1-},k_{3-}-k_{2-})- e^{-i (\Phi (k_{1-})+\Phi (k_{2-})+\Phi (k_{3-}))} \theta (k_{1-}-k_{3-},k_{3-}-k_{2-})
\\
\nonumber
&- e^{+i (\Phi (k_{1-})+\Phi (k_{2-})-\Phi (k_{3+}))} \theta (k_{1-}-k_{2-},k_{3+}-k_{1-})+ e^{+i (\Phi (k_{1-})+\Phi (k_{2-})-\Phi (k_{3+}))} \theta (k_{2-}-k_{1-},k_{1-}-k_{3+})
\\
\nonumber
&- e^{+i (\Phi (k_{1-})+\Phi (k_{2-})-\Phi (k_{3+}))} \theta (k_{2-}-k_{1-},k_{3+}-k_{2-})+e^{+i (\Phi (k_{1-})+\Phi (k_{2-})-\Phi (k_{3+}))} \theta (k_{2-}-k_{3+},k_{3+}-k_{1-})
\\
\nonumber
&+ e^{+i (\Phi (k_{1-})-\Phi (k_{2+})-\Phi (k_{3+}))} \theta (k_{1-}-k_{2+},k_{2+}-k_{3+})- e^{+i (\Phi (k_{1-})-\Phi (k_{2+})-\Phi (k_{3+}))} \theta (k_{2+}-k_{1-},k_{1-}-k_{3+})
\\
\nonumber
&+ e^{+i (\Phi (k_{1-})-\Phi (k_{2+})-\Phi (k_{3+}))} \theta (k_{1-}-k_{3+},k_{3+}-k_{2+})
- e^{+i (\Phi (k_{1-})-\Phi (k_{2+})-\Phi (k_{3+}))} \theta (k_{2+}-k_{3+},k_{3+}-k_{1-})
\\
\nonumber
&+ e^{+i (\Phi (k_{1+})-\Phi (k_{2-})-\Phi (k_{3-}))} \theta (k_{1+}-k_{2-},k_{2-}-k_{3-})- e^{+i (\Phi (k_{1+})-\Phi (k_{2-})-\Phi (k_{3-}))} \theta (k_{2-}-k_{1+},k_{1+}-k_{3-})
\\
\nonumber
&+ e^{+i (\Phi (k_{1+})-\Phi (k_{2-})-\Phi (k_{3-}))} \theta (k_{1+}-k_{3-},k_{3-}-k_{2-})- e^{+i (\Phi (k_{1+})-\Phi (k_{2-})-\Phi (k_{3-}))} \theta (k_{2-}-k_{3-},k_{3-}-k_{1+})
\\
\nonumber
&- e^{+i (\Phi (k_{1+})+\Phi (k_{2+})-\Phi (k_{3-}))} \theta (k_{1+}-k_{2+},k_{3-}-k_{1+})+ e^{+i (\Phi (k_{1+})+\Phi (k_{2+})-\Phi (k_{3-}))} \theta (k_{2+}-k_{1+},k_{1+}-k_{3-})
\\
\nonumber
&- e^{+i (\Phi (k_{1+})+\Phi (k_{2+})-\Phi (k_{3-}))} \theta (k_{2+}-k_{1+},k_{3-}-k_{2+})+ e^{+i (\Phi (k_{1+})+\Phi (k_{2+})-\Phi (k_{3-}))} \theta (k_{2+}-k_{3-},k_{3-}-k_{1+})
\\
\nonumber
&- e^{-i (\Phi (k_{1+})+\Phi (k_{2+})+\Phi (k_{3+}))} \theta (k_{1+}-k_{2+},k_{2+}-k_{3+})+ e^{+i (\Phi (k_{1+})+\Phi (k_{2+})+\Phi (k_{3+}))} \theta (k_{1+}-k_{2+},k_{3+}-k_{1+})
\\
&+ e^{+i (\Phi (k_{1+})+\Phi (k_{2+})+\Phi (k_{3+}))} \theta (k_{2+}-k_{1+},k_{3+}-k_{2+})- e^{-i (\Phi (k_{1+})+\Phi (k_{2+})+\Phi (k_{3+}))} \theta (k_{1+}-k_{3+},k_{3+}-k_{2+})
\label{I-3Phi-c:eq}
\end{align}
and
\begin{align}
\nonumber
  {\cal I}_{3\Phi(d)}&=
  \\
  \nonumber
 &-e^{+i (\Phi (k_{1+})+\Phi (k_{2+})+\Phi (k_{3+}))} \theta (k_{2-}-k_{1-},k_{3-}-k_{2-})-e^{+i (\Phi (k_{1-})+\Phi (k_{2-})+\Phi (k_{3-}))} \theta (k_{2-}-k_{1-},k_{3-}-k_{2-})
  \\
\nonumber
& +e^{-i (\Phi (k_{1-})+\Phi (k_{2-})+\Phi (k_{3-}))} \theta (k_{1-}-k_{2-},k_{1-}-k_{3-})-e^{+i (\Phi (k_{1-})+\Phi (k_{2-})-\Phi (k_{3+}))} \theta (k_{2-}-k_{1-},k_{2-}-k_{3+})
\\
\nonumber
&+e^{+i (\Phi (k_{1-})+\Phi (k_{2-})-\Phi (k_{3+}))} \theta (k_{3+}-k_{1-},k_{3+}-k_{2-})-e^{+i (\Phi (k_{1-})-\Phi (k_{2+})-\Phi (k_{3+}))} \theta (k_{1-}-k_{2+},k_{1-}-k_{3+})
\\
\nonumber
&+e^{+i (\Phi (k_{1-})-\Phi (k_{2+})-\Phi (k_{3+}))} \theta (k_{2+}-k_{1-},k_{2+}-k_{3+})-e^{+i (\Phi (k_{1+})-\Phi (k_{2-})-\Phi (k_{3-}))} \theta (k_{1+}-k_{2-},k_{1+}-k_{3-})
\\
\nonumber
&+e^{+i (\Phi (k_{1+})-\Phi (k_{2-})-\Phi (k_{3-}))} \theta (k_{2-}-k_{1+},k_{2-}-k_{3-})-e^{+i (\Phi (k_{1+})+\Phi (k_{2+})-\Phi (k_{3-}))} \theta (k_{2+}-k_{1+},k_{2+}-k_{3-})
\\
&+e^{+i (\Phi (k_{1+})+\Phi (k_{2+})-\Phi (k_{3-}))} \theta (k_{3-}-k_{1+},k_{3-}-k_{2+})+e^{-i (\Phi (k_{1+})+\Phi (k_{2+})+\Phi (k_{3+}))} \theta (k_{1+}-k_{2+},k_{1+}-k_{3+})
\label{I-3Phi-d:eq}
\end{align}
where $\theta(x,y)$ is two-dimensional Heaviside theta function which is $1$ only when both $x$ and $y$ are positive and $\Phi(k)$ is a shorthand for $\Phi(k,\alpha)$.

Now, let us make a crucial remark before moving on further. Assuming that all unit step functions $\theta$ are equal to $1$, ${\cal I}_{3\Phi(c)}$ and ${\cal I}_{3\Phi(d)}$ cancel each other. Actually, it is impossible to analytically compute the momentum integral of the above expressions. They can however be estimated using the stationary phase approximation. Hence, we should first find momentum $\p$ s for which  total phase of each term has a saddle point. The above lengthy result can be further simplified if one finds a way to get rid of step functions. Quite interestingly, we find that all step functions are are equal to $1$ in vicinity of associated saddle point.\footnote{Mathematica notebook to check this assertion is available \href{http://abolhasani.physics.sharif.edu/files/calcs/Total-Cancellation-Mom_Integration.nb}{here}.} Henceforth, one can deduce that three-point function of scalar perturbations does not get any contribution from diagrams with negligible energy transfer at all vertices.

\section{Two-point function, two-loop calculation}\label{appen:two-loop-2pt-funcn}

As a further check for our general discussions in section \ref{sec:6}, here we present an explicit two-loop two-point function analysis.
As is depicted in Fig. \ref{Types-2-loop-diagrams} there are generally three different \emph{types} of diagrams contributing to two-point function at two loops.
Two-loop diagrams have four interaction vertices and we need to focus on \eqref{two-point:N} with $V=4$.  Using the general in-in formulation developed in section \ref{sec:3-2}, for the two-point function of the inflaton fluctuations at two-loop level we have
\begin{align}\label{2-pt-two-loop-1}
\hspace{-5mm}\expect{\vphi(\x_1)\vphi(\x_2)}_{\mathrm{2-loop}} =\expect{\vphi_2(\x_1)\vphi_2(\x_2)}+\expect{\vphi_1(\x_1)\vphi_3(\x_2)}+\expect{\vphi_0(\x_1)\vphi_4(\x_2)}
\end{align}
In the following, without going too much to the details, we discuss which diagrams can saturate the condition $N_{E,L} \leq 3-E$,  \emph{cf.} \eqref{general-estimate-improved}.


\paragraph{$\expect{\vphi_2(\k) \vphi_2(-\k)}$.}
Using Feynman rules presented in  section \ref{sec:3-2} we have 
 \begin{align}
 \nonumber
 \expect{\vphi_2(x_1) \vphi_2(x_2)}_{22} =  \int  \prod_{i=1}^4\ d{\cal T}_i ~ \int \prod_{i=1}^4 d^3 y_i ~ &G_{\vphi}(x_1,y_1) G_{\vphi}(x_2,y_2) G_{\chi}(y_1,y_3) G_{\chi}(y_2,y_4)
 \\
 & \Big \langle\{\chi(y_1), \vphi(y_3)\chi(y_3)\}\, \{\chi(y_2), \vphi(y_4)\chi(y_4)\}\Big \rangle
 \end{align}
 where $\{A,B\}=AB+BA$ denotes the symmetrized product of corresponding operators.
 \begin{figure}[ht]
    \begin{axopicture}(70,230)(-50,-180)
    \SetWidth{1.5}
    \SetColor{Black}
    \Vertex(0,0){3.5}
    \ArrowLine(45,0)(0,0)
    \SetColor{Red}
    \ArrowArc(82.5,0)(37.5,90,180)
    \DashCArc(82.5,0)(37.5,180,270){3}
    \ArrowArc(82.5,0)(37.5,270,0) \DashCArc(82.5,0)(37.5,0,90){3}
    \SetColor{Black}
    \DashLine(82.5,-37.5)(82.5,37.5){3}
    \ArrowLine(120,0)(165,0)
    \Vertex(165,0){3.5}
    \Text(52,0){1}
    \Text(110,0){2}
    \Text(82.5,46){3}
    \Text(82.5,-46){4}
    \Text(82.5,0){\Cross}
    \Text(82.5,-60){(a)}
    \SetColor{Red}
  \Text(55,-25){\Plus}
  \Text(110,+25){\Plus}
  \end{axopicture}
  \begin{axopicture}(70,230)(-200,-180)
    \SetWidth{1.5}
    \SetColor{Black}
    \Vertex(0,0){3.5}
    \ArrowLine(45,0)(0,0)
    \SetColor{Red}
    \DashCArc(82.5,0)(37.5,0,180){3}
    \SetColor{Red}
    \ArrowArc(82.5,0)(37.5,295,360) \SetColor{Red}
    \ArrowArcn(82.5,0)(37.5,245,180)
    \SetColor{Red}
    \DashCArc(82.5,-30)(15,0,180){3} 
    \SetColor{Black}
    \DashCArc(82.5,-30)(15,180,0){3}
    \SetColor{Black}
    \ArrowLine(120,0)(165,0)
    \Vertex(165,0){3.5}
    \Text(80,-60){(b)}
    \Text(52,0){1}
    \Text(110,0){2}
    \Text(74,-32){3}
    \Text(90,-32){4}
    \Text(82.5,-45){\Cross}
    \SetColor{Red}
  \Text(82.5,37.5){\Cross}
  \Text(82.5,-15.25){\Cross}
  \end{axopicture}
  \begin{axopicture}(70,230)(40,-60)
    \SetWidth{1.5}
    \SetColor{Black}
    \Vertex(0,0){3.5}
    \ArrowLine(45,0)(0,0)
    \SetColor{Red}
    \ArrowArc(82.5,0)(37.5,0,180)
    \DashCArc(82.5,0)(37.5,180,360){3}
    \ArrowArcn(202.5,0)(37.5,180,0)
    \DashCArc(202.5,0)(37.5,180,360){3}
    \SetColor{Black}
    \DashLine(165,0)(120,0){3}
    \ArrowLine(240,0)(285,0)
    \Vertex(285,0){3.5}
    \Text(52,0){1}
    \Text(110,0){3}
    \Text(173,0){4}
    \Text(230,0){2}
    \Text(140,-60){(c)}
    \Text(142.5,0){\Cross}
    \SetColor{Red}
  \Text(82.5,-37.5){\Cross}
  \Text(202.5,-37.5){\Cross}
  \end{axopicture}
      \caption{Feynman diagrams associated with $\expect{\vphi_2(\k) \vphi_2(-\k)}$.} \label{two-loop-2-2}
  \end{figure}

For the overlapping (a) and the nested (b) diagrams, after some algebra we obtain,
 \begin{align}
 \nonumber
 \expect{\vphi_2(\k) \vphi_2(-\k)}_{22\,(a)} = \int d^3p\, d^3q\,\int  \prod_{i=1}^4 d{\cal T}_i\,& ~ G_{\vphi}(k,\tau_1) G_{\vphi}(k,\tau_2) G_{\chi}(p;\tau_1,\tau_3) G_{\chi}(p;\tau_2,\tau_4)
 \\
 & \chi_{\-\p+\k}(\tau_1) \vphi_{\q}(\tau_3)\chi_{\p-\q}(\tau_3) \chi^{\ast}_{\p-\k}(\tau_4) \vphi_{-\q}^{\ast}(\tau_2)\chi^{\ast}_{\q-\p}(\tau_4), 
 \end{align}
 \begin{align}
 \nonumber
 \expect{\vphi_2(\k) \vphi_2(-\k)}_{22\,(b)} = \int d^3p\, d^3q\,\int  \prod_{i=1}^4 d{\cal T}_i\,& ~ G_{\vphi}(k,\tau_1) G_{\vphi}(k,\tau_2) G_{\chi}(p;\tau_1,\tau_3) G_{\chi}(p;\tau_2,\tau_4)
 \\
 & \chi_{\-\p+\k}(\tau_1) \vphi_{\q}(\tau_3)\chi_{\p-\q}(\tau_3) \chi^{\ast}_{\p-\k}(\tau_2) \vphi_{-\q}^{\ast}(\tau_4)\chi^{\ast}_{\q-\p}(\tau_4) 
 \end{align} 
First, we note that these two diagrams are mirror symmetric. It is hence guaranteed that vertices 1 and 2 as well as 3 and 4 can potentially resonate in the loop momentum integral. Consequently, for $p,q \gg k$ we obtain
 \begin{align}
 \nonumber
 \dfrac{\expect{\vphi_2(\k) \vphi_2(-\k)}_{22(b)-22(a)}}{\expect{\vphi_2(\k) \vphi_2(-\k)}_{\mathrm{1-loop}}} \lesssim \left(\dfrac{{\cal G}^2 f^2 }{H^2 \alpha}\right) \frac{1}{\alpha}\int^{p\sim\alpha k} \dfrac{d^3p}{p^3}\int^{q\sim\alpha k} \dfrac{ d^3q}{q|\p-\q|}\sim \left(\frac{\mathbb{H}}{H}\right)^2 \frac{\ln\alpha}{\alpha}.
 \end{align}
We have used $\lesssim$ as we have dropped the momentum dependent phase factors which may lead to extra suppression factors.
Therefore, 
$\expect{\vphi_2(\k) \vphi_2(-\k)}_{22(a)}, \expect{\vphi_2(\k) \vphi_2(-\k)}_{22(b)}$ do not saturate the bound on the superficial degree of divergence in  \eqref{general-estimate-improved}.
 
Moreover, for the last one diagram (c) we have learned from the one-loop  two-point function calculations  that a loop with a dashed and a solid lines is suppressed by an extra factor of $\alpha^{-1}$ so the associated momentum integral of of  $\expect{\vphi_2(\k) \vphi_2(-\k)}_{22(c)}$ is at most of the order of $\sim (\log \alpha /\alpha^2)$.

  \paragraph{{$\expect{\vphi_3(\k) \vphi_1(-\k)}$}.}
The associated Feynman graphs are drawn below.
   \begin{figure}[H]
  \begin{axopicture}(70,210)(-50,-180)
    \SetWidth{1.5}
    \SetColor{Black}
    \Vertex(0,0){3.5}
    \ArrowLine(45,0)(0,0)
    \SetColor{Red}
    \DashCArc(82.5,0)(37.5,0,180){3}
    \SetColor{Red}
    \DashCArc(82.5,0)(37.5,295,360){3} \SetColor{Red}
    \ArrowArcn(82.5,0)(37.5,245,180)
    \SetColor{Red}
    \ArrowArc(82.5,-34)(15,0,180) 
    \SetColor{Black}
    \DashCArc(82.5,-34)(15,180,0){3}
    \SetColor{Black}
    \ArrowLine(120,0)(165,0)
    \Vertex(165,0){3.5}
    \Text(80,-65){(a)}
  \Text(82.5,-49){\Cross}
    \SetColor{Red}
    \Text(82.5,37.5){\Cross}
  \Text(114.0,-20.25){\Plus}
  \end{axopicture}
  \begin{axopicture}(70,230)(-200,-180)
    \SetWidth{1.5}
    \SetColor{Black}
    \Vertex(0,0){3.5}
    \ArrowLine(45,0)(0,0)
    \SetColor{Red}
    \DashCArc(82.5,0)(37.5,0,180){3}
    \SetColor{Red}
    \DashCArc(82.5,0)(37.5,295,360){3} \SetColor{Red}
    \ArrowArcn(82.5,0)(37.5,245,180)
    \SetColor{Black}
    \ArrowArc(82.5,-34)(15,0,180) 
    \SetColor{Red}
    \DashCArc(82.5,-34)(15,180,0){3}
    \SetColor{Black}
    \ArrowLine(120,0)(165,0)
    \Vertex(165,0){3.5}
    \Text(80,-65){(b)}
    \SetColor{Red}
    \Text(82.5,-49){\Cross}
  \Text(82.5,37.5){\Cross}
  \Text(114.0,-20.25){\Plus}
  \end{axopicture}
  \begin{axopicture}(70,230
 )(120,-60)
    \SetWidth{1.5}
    \SetColor{Black}
    \Vertex(0,0){3.5}
    \SetColor{Black}
    \ArrowLine(45,0)(0,0)
    \SetColor{Red}
    \ArrowArc(82.5,0)(37.5,90,180)
    \SetColor{Red}
    \ArrowArcn(82.5,0)(37.5,270,180)
    \SetColor{Red}
    \DashCArc(82.5,0)(37.5,270,90){3}
    \SetColor{Black}
    \DashLine(82.5,-37.5)(82.5,37.5){3}
    \SetColor{Black}
    \ArrowLine(120,0)(165,0)
    \Vertex(165,0){3.5}
    \Text(80,-60){(c)}
    \Text(82.5,0){\Cross}
    \SetColor{Red}
 \Text(110,+25){\Plus}
  \Text(110,-25){\Plus}
  \end{axopicture}
  \begin{axopicture}(70,230)(0,-60)
    \SetWidth{1.5}
    \SetColor{Black}
    \Vertex(0,0){3.5}
    \ArrowLine(45,0)(0,0)
    \SetColor{Red}
    \ArrowArc(82.5,0)(37.5,0,180)
    \DashCArc(82.5,0)(37.5,180,360){3}
    \DashCArc(202.5,0)(37.5,0,180){3}
    \DashCArc(202.5,0)(37.5,180,360){3}
    \SetColor{Black}
    \ArrowLine(165,0)(120,0)
    \ArrowLine(240,0)(285,0)
    \Vertex(285,0){3.5}
    \Text(140,-60){(d)}
    \SetColor{Red}
\Text(82.5,-37.5){\Cross}
  \Text(202.5,37.5){\Cross}
  \Text(202.5,-37.5){\Cross}    
  \end{axopicture}
  \caption{Feynman diagrams associated with $\expect{\vphi_3(\k) \vphi_1(-\k)}$.}\label{two-loops-1-3}
  \end{figure}
Without delving into details, employing what we have learned so far, we can infer that the nested diagrams (a) and (b) do not saturate the $\alpha$ dependence of momentum integrals, as they are  made up of mixed loops of solid and dashed lines. However, diagram (c) can potentially saturate that condition. Finally diagram (d) has two sources of suppression, first the left loop is a mixed loop which cause a $1/\alpha$ suppression, second the connecting propagator is vanishing in the limit where two loops are at the resonance.
  
  \paragraph{$\expect{\vphi_0(\k) \vphi_4(-\k)}$.} There are large number of diagrams contributing to this correlator, some of which are depicted below.
  \vskip 7mm
\begin{figure}[h]
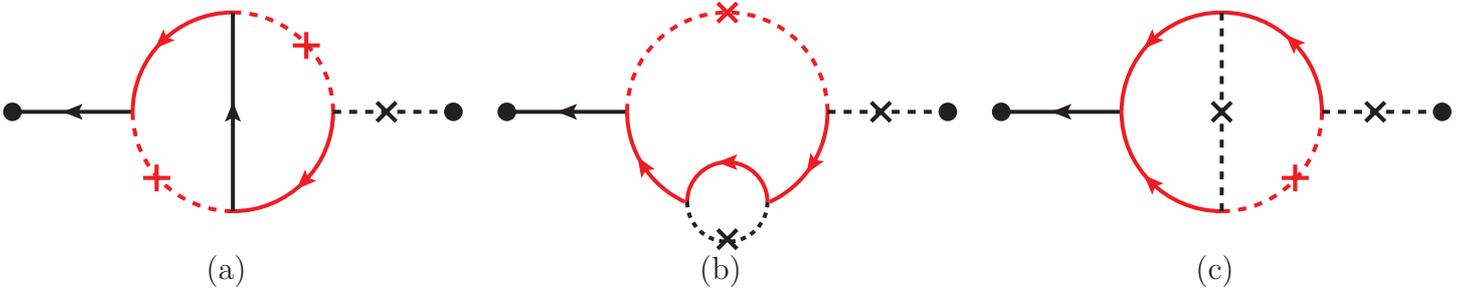

\begin{axopicture}(70,80)(10,-60) 
    \SetWidth{1.5}
    \SetColor{Black}
    \Vertex(0,0){3.5}
    \ArrowLine(45,0)(0,0)
    \SetColor{Red}
    \ArrowArc(82.5,0)(37.5,90,180)
    \SetColor{Red}
    \ArrowArcn(82.5,0)(37.5,0,270)
    \SetColor{Red}
    \DashCArc(82.5,0)(37.5,0,90){3}
    \SetColor{Red}
    \DashCArc(82.5,0)(37.5,180,270){3}
    \SetColor{Black}
    \ArrowLine(82.5,-37.5)(82.5,37.5)
    \DashLine(120,0)(165,0){3}
    \Text(80,-60){(a)}
    \Vertex(165,0){3.5}
  \Text(140,0){\Cross}
    \SetColor{Red}
 \Text(110,+25){\Plus}
  \Text(54,-25){\Plus}
  \end{axopicture}
  \begin{axopicture}(70,80)(-105,-60)
    \SetWidth{1.5}
    \SetColor{Black}
    \Vertex(0,0){3.5}
    \ArrowLine(45,0)(0,0)
    \SetColor{Red}
    \DashCArc(82.5,0)(37.5,0,180){3}
    \SetColor{Red}
    \ArrowArcn(82.5,0)(37.5,245,180)
    \SetColor{Red}
    \ArrowArcn(82.5,0)(37.5,0,295)
    \SetColor{Red}
    \ArrowArc(82.5,-34)(15,0,180) 
    \SetColor{Black}
    \DashCArc(82.5,-34)(15,180,0){2}
    \SetColor{Black}
    \DashLine(120,0)(165,0){3}
    \Vertex(165,0){3.5}
        \Text(80,-60){(b)}
        \Text(140,0){\Cross}
        \Text(82.5,-48){\Cross}
    \SetColor{Red}
    \Text(82.5,37.5){\Cross}
  \end{axopicture} 
  \begin{axopicture}(70,80)(-220,-60)
    \SetWidth{1.5}
    \SetColor{Black}
    \Vertex(0,0){3.5}
    \ArrowLine(45,0)(0,0)
    \SetColor{Red}
    \ArrowArc(82.5,0)(37.5,0,90)\ArrowArc(82.5,0)(37.5,90,180)
    \SetColor{Red}
    \ArrowArcn(82.5,0)(37.5,270,180)
    \SetColor{Red}
    \DashCArc(82.5,0)(37.5,270,0){3}
    \SetColor{Black}
    \DashLine(82.5,-37.5)(82.5,37.5){3}
    \DashLine(120,0)(165,0){3}
    \Text(80,-60){(c)}
    \Vertex(165,0){3.5}
    \Text(82.5,0){\Cross}
    \Text(140,0){\Cross}
     \SetColor{Red}
 \Text(110,-25){\Plus}
  \end{axopicture}
  \caption{Sample Feynman diagrams associated with $\expect{\vphi_0(\k) \vphi_4(-\k)}$. Solid lines depict the retarded Green function while dashed lines are contraction of two fields; the red and black colors respectively correspond to $\chi, \vphi$ fields.}\label{04-Diagrams}
  \end{figure}
With the same reasoning as in the previous section one can investigate which diagram gives the dominant $\alpha$ dependence to the momentum integrals. For example, one can show that like diagram (a) of Fig. \ref{two-loop-2-2}, pairs of vertices facing each other in diagrams (a) and (c) above can resonate in momentum integral and saturate the condition of \eqref{general-estimate-improved}. However, since the diagram (c) has a mixed loop, as discussed before, it is  suppressed at least by an extra $1/\alpha$ factor so does not saturate the $\alpha$-power expected from superficial power-counting in the loop momentum integral.

 

\section{General structure of $L$-loop diagrams}\label{appen:loop-insertion}

For our momentum localization/cutoff arguments in section \ref{sec:6} we need to discuss some basic  features of $L$-loop diagrams. Since in the large momentum limit both $\vphi$ and $\chi$ can be considered as massless fields, it is not crucial to differentiate between $\vphi, \chi$ fields. So, to avoid cluttering we may drop the color-coding in the Feynman graphs. Nonetheless it is  important to discriminate between contractions and propagators. 
As before, we would be only considering connected diagrams, but they can now be 1-PI or non 1-PI.

\begin{figure}[ht]
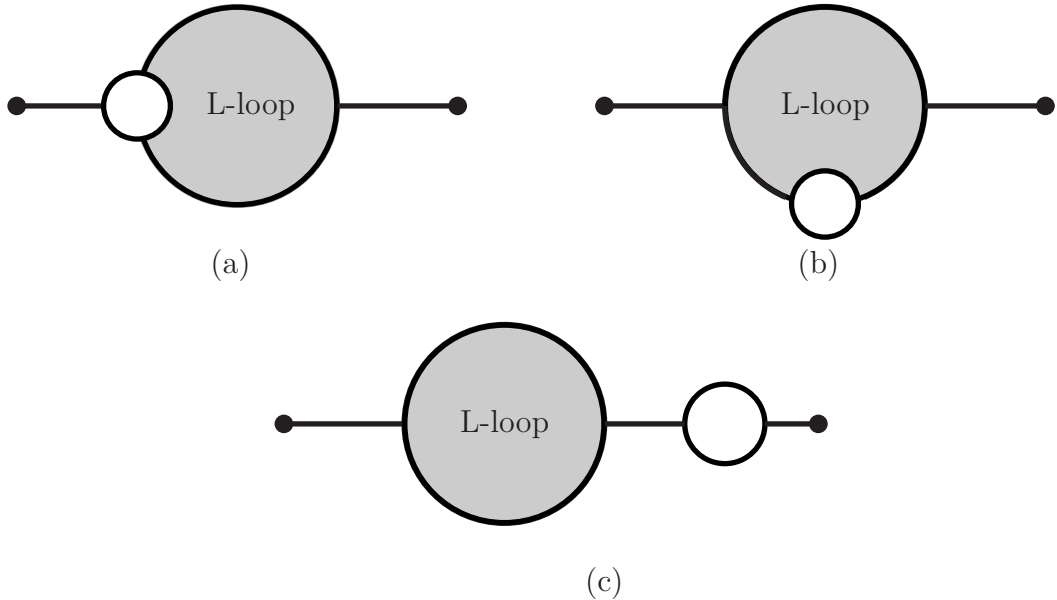

    \begin{axopicture}(70,210)(-50,-180)
    \SetWidth{2}
    \SetColor{Black}
    \Vertex(0,0){3.5}
    \Line(45,0)(0,0)
    \SetColor{Gray}
    \Arc(82.5,0)(37.5,90,160)
    \Arc(82.5,0)(37.5,200,270)
    \Arc(82.5,0)(37.5,270,0) \Arc(82.5,0)(37.5,0,90)
    \GCirc(82.5,0){37.25}{0.8}
   \SetColor{Black}
    \GCirc(45,0){12.5}{1}
    \Text(87.5,0){L-loop}
    \Line(120,0)(165,0)
    \Vertex(165,0){3.5}
    \Text(80,-60){(a)}
  \end{axopicture}
  \begin{axopicture}(70,210)(-200,-180)
    \SetWidth{2}
    \SetColor{Black}
    \Vertex(0,0){3.5}
    \Line(45,0)(0,0)
    \Arc(82.5,0)(37.5,0,180)
    \Arc(82.5,0)(37.5,293,360)  
    \GCirc(82.5,0){37.25}{0.8}
    \Text(82.5,0){L-loop}
    \Arc(82.5,0)(37.5,180,247)
   \GCirc(82.5,-37){12.5}{1}
    \Line(120,0)(165,0)
    \Vertex(165,0){3.5}
    \Text(80,-60){(b)}
    \end{axopicture}
  \begin{axopicture}(70,210)(-10,-60)
    \SetWidth{2}
    \SetColor{Black}
    \Vertex(0,0){3.5}
    \Line(45,0)(0,0)
    \Arc(82.5,0)(37.5,0,180)
    \Arc(82.5,0)(37.5,180,360) \GCirc(82.5,0){37.25}{0.8}\Text(82.5,0){L-loop}
    \GCirc(165,0){15}{1}
    \Line(120,0)(150,0)
    \Line(180,0)(200,0)
    \Vertex(200,0){3.5}
    \Text(120,-60){(c)}
  \end{axopicture}
      \caption{General structure of $L+1-$loop two-point function diagrams: (a) overlapping (b) nested and (c) non 1-PI diagrams. The above  shows the type of,  and  not the actual, Feynman graphs.}\label{Types-2-loop-diagrams}
  \end{figure}

\paragraph{Generic higher loop diagrams.} An $L+1$-loop diagram can be constructed by inserting a new loop either into a line (dashed or solid line) or a vertex of a given $L$-loop diagram. Let us start with a one-loop two-point function. There are three possible ways of doing so, as depicted in Fig. \ref{Types-2-loop-diagrams}.
One may readily observe that the one and two-loop two-point function  and the one-loop three-point function diagrams discussed earlier illustrate special examples of the above general construction.

\end{document}